\documentclass[prd,12pt]{article}
\pdfoutput=1

\usepackage{amsmath,amssymb,graphicx,multirow,xspace,slashed}
\usepackage[colorlinks=true,urlcolor=blue,anchorcolor=blue,citecolor=blue,filecolor=blue,linkcolor=blue,menucolor=blue,pagecolor=blue]{hyperref}
\usepackage[compress,numbers]{natbib}
\usepackage{subfigure, placeins}
\usepackage{booktabs}
\usepackage{blindtext, rotating}
\usepackage{afterpage}
\usepackage{enumitem}
\usepackage{ marvosym }
\usepackage{authblk} 
\usepackage{verbatim}
\usepackage{soul} 
\usepackage[normalem]{ulem}
\usepackage{pifont}

\usepackage[utf8]{inputenc}

\usepackage{floatrow}
\newfloatcommand{capbtabbox}{table}[][\FBwidth]

\usepackage[font=footnotesize,labelfont=bf]{caption}

\usepackage{lineno}

\allowdisplaybreaks

\long\def\symbolfootnote[#1]#2{\begingroup%
\def\thefootnote{\fnsymbol{footnote}}\footnote[#1]{#2}\endgroup}

\newcommand{\newc}{\newcommand}
\newc{\gsim}{\lower.7ex\hbox{$\;\stackrel{\textstyle>}{\sim}\;$}}
\newc{\lsim}{\lower.7ex\hbox{$\;\stackrel{\textstyle<}{\sim}\;$}}
\newc{\gev}{\,{\rm GeV}}
\newc{\mev}{\,{\rm MeV}}
\newc{\ev}{\,{\rm eV}}
\newc{\kev}{\,{\rm keV}}
\newc{\tev}{\,{\rm TeV}}
\newc{\MHT}{$H_T^{\text{miss}}$}
\newc{\MET}{$\slashed{E}_T$}
\newc{\MTT}{$M_{T2}$}

\newc{\mz}{M_Z}
\newc{\mpl}{M_*}
\newc{\mw}{m_{\rm weak}}
\newc{\nr}[1]{N^c_R{}_{#1}}

\def\beq{\begin{equation}}
\def\eeq{\end{equation}}
\newcommand{\bea}{\begin{eqnarray}\begin{aligned}}
\newcommand{\eea}{\end{aligned}\end{eqnarray}}
\def\bitem{\begin{itemize}}
\def\eitem{\end{itemize}}

\numberwithin{equation}{section}

\newcommand\fverb{\setbox\fverbbox=\hbox\bgroup\verb}

\newbox\fverbbox

\usepackage[usenames,dvipsnames]{xcolor}

\begin{document}

\baselineskip 0.6cm

\begin{titlepage}

\thispagestyle{empty}

\begin{center}

\vskip 1cm

{\Large \bf It's all right(-handed neutrinos): }\vskip0.3cm
{\Large \bf a new $W'$ model for the $R_{D^{(*)}}$  anomaly}

\vskip 1.0cm
{\large Pouya Asadi$^{1,2}$, Matthew~R.~Buckley$^1$, and David Shih$^1$ }
\vskip 1.0cm
{\it $^1$ NHETC, Dept.~of Physics and Astronomy\\ Rutgers, The State University of NJ \\ Piscataway, NJ 08854 USA} \\
\vskip 0.5cm
{\it $^2$ Kavli Institute for Theoretical Physics\\ University of California\\ Santa Barbara, CA 93106, USA} \\
\vskip 1.0cm

\end{center}

\begin{abstract}
The measured $B$-meson semi-leptonic branching ratios $R_{D}$ and $R_{D^*}$ have long-standing deviations between theory and experiment. We introduce a model which explains both anomalies through a single interaction by introducing a right-handed neutrino as the missing energy particle. This interaction is mediated by a heavy charged vector boson ($W'$) which couples only to right-handed quarks and leptons of the Standard Model through  the mixing of these particles with new vector-like fermions. Previous $W'$ models for the $R_{D^{(*)}}$ anomaly  were strongly constrained from flavor changing neutral currents and direct collider searches for $Z'\to\tau\tau$ resonances. 
We show that relying on right-handed fermion mixing enables us to avoid these constraints, as well as other severe bounds from electroweak precision tests and neutrino mixing.
\end{abstract}

\vskip5cm

\center{{\it Dedicated to the memory of Diane Soyak}}

\flushbottom

\end{titlepage}

\setcounter{page}{1}

\tableofcontents

\vfill\eject

\section{Introduction}
\label{sec:intro}

Experimental tests of the Standard Model (SM) have probed many different aspects of potential new physics (NP), 
including direct searches for new heavy particles at the Large Hadron Collider (LHC), various direct and indirect dark matter detection experiments, neutrino experiments, and precision measurements of flavor physics. For the most part, predictions from the SM are in good agreement with the results from these experiments. There are, however, a handful of anomalies which suggest the existence of new physics.  

Arguably, some of the most significant and enduring discrepancies with SM predictions are observed in $B$-physics experiments.  Collaborations such as BaBar \cite{Aubert:2007dsa,Lees:2012xj, Lees:2013uzd}, Belle \cite{Bozek:2010xy,Huschle:2015rga, Abdesselam:2016xqt}, and LHCb \cite{Aaij:2015yra,Aaij:2017tyk,Aaij:2017vbb}, have observed anomalies in the rate of $B$-hadron decays, compared to the theoretical predictions of the SM. The most significant deviations from the SM predictions are found in the semi-leptonic decay of $B$ mesons to $D$ or $D^*$, encapsulated in the ratios $R_D$ and $R_{D^{*}}$, defined as 
\begin{equation}
R_D = \frac{\Gamma (\bar{B} \rightarrow D \tau \nu)}{\Gamma (\bar{B} \rightarrow D \ell \nu)}, \hspace{0.4in}  R_{D^{*}} = \frac{\Gamma (\bar{B} \rightarrow D^{*} \tau \nu)}{\Gamma (\bar{B} \rightarrow D^{*} \ell \nu)},
\label{eq:rddef}
\end{equation}
where $\ell$ stands for either electrons or muons. The global average \cite{Amhis:2016xyh} of the observed values 
\cite{Aubert:2007dsa,Lees:2012xj, Lees:2013uzd, Huschle:2015rga, Abdesselam:2016xqt,Aaij:2015yra} 
 for these ratios are 
\begin{equation}
R_D = 0.403 \pm 0.040 \pm 0.024 , \hspace{0.5 in}  R_{D^{*}} = 0.310 \pm 0.015 \pm 0.008,
\label{eq:rdobs}
\end{equation}
where the first (second) experimental errors are due to statistics (systematics). 
Meanwhile the Standard Model predictions for these ratios are \cite{Lees:2012xj,Lees:2013uzd,Amhis:2016xyh,Aoki:2016frl,Na:2015kha,Fajfer:2012vx,Lattice:2015rga,Bailey:2012jg,Kamenik:2008tj,Jaiswal:2017rve}
\begin{equation}
R_D =0.300 \pm 0.008 , \hspace{0.5 in}  R_{D^{*}} = 0.252 \pm 0.003,
\label{eq:rdsm}
\end{equation}
which is in sharp disagreement with the experimental values reported by different collaborations.  A combined analysis \cite{Amhis:2016xyh} shows a $\sim 3.9\sigma$ discrepancy with the SM predictions of Eq.~\eqref{eq:rdsm}. 
It is proposed \cite{Kim:2016yth} that $\sim 10\%$ modification of some form-factors obtained through lattice calculation can slightly reduce the discrepancy with SM prediction of one of these anomalies ($R_D$). Nonetheless, given the large deviation between the SM predictions and the observed values, an investigation of different possible theoretical explanations beyond the SM is well-motivated.

Many theoretical models have been put forward to explain the $R_D$ and $R_{D^*}$ anomalies. Given that the measured $R_{D^{(*)}}$ ratios are higher than their SM predictions, model building efforts have focused on enhancing the rate of $b\to c\tau\nu$ transitions through new mediating particles (this is much easier than suppressing the rate of $b\to c(e,\mu)\nu$ transitions, given the much more stringent constraints on new physics coupling to electrons and muons). Integrating out the heavy mediators along with the $W$ at tree-level results in a dimension-6 effective Hamiltonian of the form
\beq\label{Heff}
 {\mathcal H}_{\rm eff} = \frac{4 G_F V_{cb}}{\sqrt{2}}  \left( {\mathcal O}^V_{LL} + \sum_{X=S,V,T\atop M,N=L,R} C^X_{MN}{\mathcal O}^X_{MN} \right)
\eeq
where the four-fermion effective operators are defined as
\begin{eqnarray}
 {\mathcal O}^S_{MN} & \equiv & (\bar c P_M b)(\bar \tau P_N \nu) \nonumber \\
{\mathcal O}^V_{MN} & \equiv &(\bar c \gamma^\mu P_M b)(\bar \tau \gamma_\mu P_N \nu) \label{eq:Lop}\\
{\mathcal O}^T_{MN} &\equiv &(\bar c \sigma^{\mu\nu} P_M b)(\bar \tau \sigma_{\mu\nu}P_N \nu), \nonumber
\end{eqnarray}
for $M,N = R$ or $L$. We have separated out the SM contribution in the first term of Eq.~\eqref{Heff}; the normalization factor is conventional and chosen so that $(C^V_{LL})_{SM}=1$. 
As we will review below, ${\mathcal O}^V_{LL}$ is unique among all the operators which can be built out of SM fields, in that it can explain both $R_D$ and $R_{D^{*}}$ simultaneously. 

\begin{figure}
\includegraphics[scale=1]{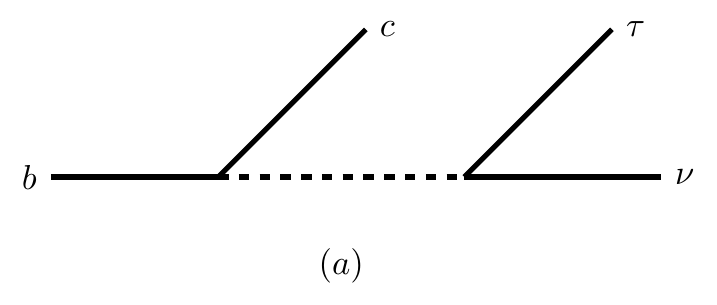}
\includegraphics[scale=1]{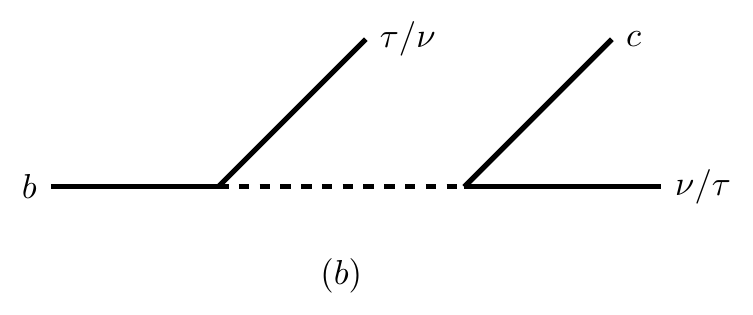}
\caption{The generic diagrams that contribute to $R_D$ and $R_{D*}$ by modifying $b \rightarrow c \tau \nu$ amplitudes with a BSM mediator. 
The mediator can be one of the three candidates indicated in the text: (a) charged Higgs or $W'$; or (b) leptoquarks.}
\label{fig:diagram}
\end{figure}

The mediators can be spin-0 or spin-1, and they can either carry baryon and lepton number (leptoquarks) or be $B/L$ neutral (charged Higgs and $W'$). 
These possibilities are illustrated in Fig.~\ref{fig:diagram}. Existing models  
can be divided into three general categories:
\begin{itemize}

\item \textit{Extended Higgs sector} \cite{Tanaka:2010se,Fajfer:2012jt,Crivellin:2012ye,Celis:2012dk}.
Integrating out a charged Higgs produces the scalar-scalar operators ${\mathcal O}^S$. 
These operators are severely constrained by the measured $B_c$ lifetime \cite{Li:2016vvp,Celis:2016azn,Alonso:2016oyd, Akeroyd:2017mhr}, 
which rules out nearly all explanations of the observed  $R_{D^{(*)}}$ using this class of models. It should be noted that these constraints are generic to all models in this category; even general extensions of the Higgs sector, for example Type-III two-Higgs-doublet models (2HDMs), are strongly disfavored for these anomalies.

\item \textit{Heavy charged vector bosons} \cite{He:2012zp,Boucenna:2016qad}. Integrating out $W'$'s gives rise to the vector-vector 
operators ${\mathcal O}^V$. To explain both $R_D$ and $R_{D^{*}}$ simultaneously with left-handed neutrinos, $C^V_{LL}$ should be non-zero. Constraints on these models arise from the inevitable existence of an accompanying $Z'$ mediator.  By $SU(2)$ invariance, the $W'b_Lc_L$ vertex implies a $Z'b_Ls_L$ vertex. In order to avoid catastrophic tree-level flavor-changing neutral currents (FCNCs) from this $Z'$, some mechanism to suppress the $Z'b_Ls_L$ vertex -- for example, minimal flavor violation (MFV) -- must be assumed \cite{Greljo:2015mma,Faroughy:2016osc}. However, this will not suppress $Z'bb$ and $Z'\tau\tau$ vertices in general. In such models, there are therefore typically severe constraints from LHC direct searches for $Z'\to\tau\tau$ resonances. Evading these limits requires one to go to unnaturally high $Z'$ widths \cite{Faroughy:2016osc,Crivellin:2017zlb}.

\item \textit{Leptoquarks} \cite{Fajfer:2012jt,Tanaka:2012nw}. Leptoquarks couple quarks and leptons at a vertex. Other than their spin (which can be either zero or one), leptoquarks can be categorized by their representation under SM gauge groups. Appropriate choices of these quantum numbers can give rise to any of the operators in Eq.~\eqref{eq:Lop} after Fierz rearrangement. 
Given the wide variety of leptoquark models, there are many potentially relevant constraints, ranging from $b\rightarrow s \nu\nu$ flavor bounds \cite{Crivellin:2017zlb}, to LHC searches for $\tau\tau$ resonances \cite{Faroughy:2016osc,Crivellin:2017zlb}, and measurements of the $B_c$ life-time \cite{Alonso:2016oyd, Akeroyd:2017mhr}.  Nevertheless, viable leptoquark models exist (with either spin-0 and spin-1), and so they are considered to be the favored explanations for the $R_{D^{(*)}}$ anomaly \cite{Crivellin:2017zlb,Calibbi:2017qbu}, because the alternatives (as described above) are even more stringently constrained.

\end{itemize}

In this paper we revisit the $W'$ models and identify a new class which avoids the pitfalls described above. All of the existing $W'$ models assume that the missing energy in the semi-leptonic $B$ decay is a SM neutrino.\footnote{RH neutrinos have been combined with leptoquarks in \cite{He:2012zp,Becirevic:2016yqi} and extended Higgs sector in \cite{Cline:2015lqp,Iguro:2018qzf}; a model-independent study has been done in \cite{Bardhan:2016uhr,Dutta:2013qaa,Dutta:2016eml}. 
} Our key modification is to make the $\nu$ enhancing the $B\to D^{(*)}\tau\nu$ rate a light {\it right}-handed neutrino, rather than a left-handed neutrino of the SM. As we will show, cosmological and astrophysical considerations require $m_{\nu_R}\lesssim 10$~keV, in which case
the kinematics of the $B$ decay into this new particle would be indistinguishable from decays to the (nearly massless) SM neutrinos. Once we integrate out the $W'$ at tree level we generate the dimension six operator $\mathcal{O}^V_{RR}$. We will show that (similar to $\mathcal{O}^V_{LL}$) this single operator can explain both $R_{D}$ and $R_{D^{*}}$ simultaneously.

Furthermore, by having the $W'$ and $Z'$ couple only to right-handed fermions in the SM (through mixing with heavy vector-like fermions charged under the extra $SU(2)$), we can couple the $W'$ directly to $c_R$ and $b_R$ (instead of to the $q_{L2}=(c_L,s_L)$ and $q_{L3}=(t_L,b_L)$) and so can avoid the $Z'bs$ vertex. Thus there is no danger of tree-level FCNCs in this model, and we  obviate the need for the $Z'$ coupling to the third generation fermions to be enhanced by $1/V_{cb}$ when compared to the $W'bc$ coupling required to explain the anomalies. 
This alleviates the stringent bounds from LHC direct searches for $Z'\to\tau\tau$ resonances which were the main obstacles to previous $W'$ models. That said, we find that these searches still set meaningful bounds on the parameter space of our $W'$ model. These can be satisfied while still keeping the model perturbative, but it requires a mild enhancement to the $Z'$ width ($\Gamma_{Z'}/m_{Z'}\sim 3-10\%$). We achieve this enhancement by introducing additional vector-like matter charged under the extra $SU(2)$ which do not mix with the SM.

The additional $SU(2)$ and the fermion mixing we introduce between new vector-like fermions and SM fields can modify the relation between $W$ and $Z$ masses and the couplings of SM fermions to $W$ and $Z$. These are constrained by electroweak precision (EWP) tests. However, the EWP constraints are much milder than in models  where the two $SU(2)$'s are broken down to the diagonal by a bifundamental vev (see, e.g.~\cite{Boucenna:2016qad, Donini:1997yu}), as there is no $W$-$W'$ mixing. Additional constraints come from the effect new right-handed light neutrinos have on the cosmic microwave background (CMB) power spectrum, as well as their  mixing with SM left-handed neutrinos. Flavor constraints such as FCNCs can be evaded by a suitable choice of fermion mixing, which eliminate FCNCs at tree-level. As we will show, our model survives all current experimental tests, while having some prospect of being discovered or ruled out by the future searches.

The outline of the paper is as follows. In Section~\ref{sec:generalremarks} we explore the numerical contribution of all Wilson coefficients from the Hamiltonian in Eq.~\eqref{Heff} to $R_{D^{(*)}}$. We show how three of them -- $C^V_{LL}$, $C^V_{RR}$ and $C^V_{LR}$ -- are special in that they can each 
single-handedly explain both $R_{D}$ and $R_{D^{*}}$ anomalies. In Section~\ref{sec:model} we introduce our model: an $SU(2)$ extension of the SM that generates $C^V_{RR}$ through the combination of a $W'$ and a RH neutrino. We calculate the spectrum and the couplings of the model, in preparation for the study of its phenomenology in Section~\ref{sec:pheno}. The phenomenological consequences of the model include: electroweak precision (EWP) tests, collider signatures, cosmology and astrophysics, and more. Using these, we determine the experimental limits on the model, and show that a robust, viable parameter space exists that can explain the $R_{D^{(*)}}$ anomalies. Finally, we conclude in Section~\ref{sec:conclusion}. More details about some couplings in our model relevant for the EWP bounds can be found in Appendix~\ref{app:details}. 

\section{General Remarks}
\label{sec:generalremarks}

In this section, we will review the contributions to  $R_D$ and $R_{D^*}$ from each of the dimension six operators in Eq.~\eqref{Heff}, and discuss how this motivates model building with $W'$'s and RH neutrinos. We begin by writing down useful and fully-general numerical formulas for $R_D$ and $R_{D^*}$ in the presence of ${\mathcal H}_{\rm eff}$:
\bea
\label{eq:allCRD}
R_D &  \approx   R_D^{SM} \times \left\lbrace \left(  |1+C^V_{LL}+C^V_{RL}|^2 + |C^V_{RR}+C^V_{LR}|^2 \right)    \right.  \\
&  +  1.35  \left( |C^S_{RL}+C^S_{LL}|^2 + |C^S_{LR}+C^S_{RR}|^2 \right)  + 0.70 	\left( |C^T_{LL}|^2 +  |C^T_{RR}|^2 \right)    \\
& +   1.72 \mathcal{R}e \left[ (1+C^V_{LL}+C^V_{RL})(C^S_{RL}+C^S_{LL})^* + (C^V_{RR}+C^V_{LR})(C^S_{LR}+C^S_{RR})^* \right]     \\
& + \left. 1.00  \mathcal{R}e \left[  (1+C^V_{LL}+C^V_{RL})(C^T_{LL})^*  +   (C^V_{LR}+C^V_{RR})(C^T_{RR})^* \right] \right\rbrace   ,\\\\
R_{D^{*}} & \approx  R_{D^*}^{SM}\times  \left\lbrace  \left( |1+C^V_{LL}|^2+|C^V_{RL}|^2	+ |C^V_{LR}|^2+|C^V_{RR}|^2	\right)  \right.  \\
& +  0.04	\left( |C^S_{RL}-C^S_{LL}|^2 +  |C^S_{LR}-C^S_{RR}|^2 \right)   \\
& +   12.11 \left(  |C^T_{LL}|^2 +  |C^T_{RR}|^2 \right)   - 1.78  \mathcal{R}e \left[ (1+C^V_{LL})(C^V_{RL})^* +  C^V_{RR} (C^V_{LR})^* \right]  \\
& + 5.71 \mathcal{R}e \left[	C^V_{RL} (C^T_{LL})^*	+  C^V_{LR} (C^T_{RR})^*	\right]   -  4.15				\mathcal{R}e \left[	(1+C^V_{LL}) (C^T_{LL})^*	+ C^V_{RR} (C^T_{RR})^*	\right]   \\
& +  \left. 0.12 \mathcal{R}e\left[	(1+C^V_{LL}-C^V_{RL}) (C^S_{RL}-C^S_{LL})^*	+  (C^V_{RR}-C^V_{LR}) (C^S_{LR}-C^S_{RR})^*\right] \right\rbrace  . 
\eea
To derive these formulas without calculating any new form factors or matrix elements, one can use the following trick: we expect that these formulas should be invariant under interchange of $R$ and $L$ (i.e.\ parity) since we sum over all polarizations and spins in the end. Thus we can start from the results in the literature for left-handed neutrinos, and map them using parity to the results for right-handed neutrinos. Since there is no interference between operators with left- and right-handed neutrinos, this mapping does not miss any contributions from mixed terms. 

The analytic formulae for the differential decay rates $d\Gamma(B\to D^{(*)}\tau\nu)/dq^2$ (using only the operators that involve the SM neutrinos) are calculated in~\cite{Sakaki:2013bfa}.\footnote{We are using a slightly different naming convention for the Wilson coefficients and operators than~\cite{Sakaki:2013bfa}. The map between our convention and the one used in~\cite{Sakaki:2013bfa} is
\bea
\label{eq:WCmap}
C^V_{LL} & \rightarrow  C_{V1},\qquad C^V_{RL}  \rightarrow  C_{V2} \\
C^S_{LL} & \rightarrow  C_{S2},\qquad  C^S_{RL}  \rightarrow  C_{S1},\\
C^T_{LL} & \rightarrow  C_{T}.  
\eea
 } 
We then integrate the differential decay rates over the momentum transfer in the interval $q^2 \in \left(m_\tau^2 , (m_B - m_{D^{(*)}})^2\right)$, and substitute the numerical values in Table~\ref{tab:numbers} for all the SM parameters \cite{Patrignani:2016xqp}. This results in the numerical expressions shown in Eq.~\eqref{eq:allCRD}. 

\begin{table*}[t]
\begin{tabular}{|c|c|c|c|}
\hline 
$V_{cb}$ & $G_F$ [GeV$^{-2}$] & $m_{\bar{B}}$ [GeV] & $m_D$ [GeV] \\ 
\hline 
$42.2 \times 10^{-3}$ & $1.166 \times 10^{-5}$ & 5.279 & 1.870  \\ 
\hline 
\hline 
$m_{D^*}$ [GeV] & $m_e$ [GeV] & $m_\mu$ [GeV] & $m_\tau$ [GeV] \\
\hline 
2.010 & $0.511 \times 10^{-3}$ & 0.106 & 1.777 \\
\hline 
\end{tabular} 
\caption{The numerical values of different variables used in deriving Eqs.~\eqref{eq:allCRD}. } 
\label{tab:numbers}
\end{table*}

We have corroborated this result by directly calculating, from scratch, the contribution of operators involving right-handed neutrinos to $R_{D^{(*)}}$. This involves first calculating the matrix element of $\bar{B}\rightarrow D^{(*)}\tau\nu_R$ using the Hamiltonian in Eq.~\eqref{Heff}. The matrix element factorizes into a leptonic side, which can be calculated straightforwardly, and a hadronic side \cite{Tanaka:2012nw, Hagiwara:1989cu}. The hadronic matrix elements are functions of the masses, the  momentum transfer, and a handful of known form factors. A list of these form factors, the leptonic matrix elements with left-handed neutrinos, and the hadronic matrix elements can be found in~\cite{Tanaka:2012nw,Bardhan:2016uhr,Sakaki:2013bfa}. Specifically, for the $\bar{B}\rightarrow D \tau \nu$ we use the same form factors as in \cite{Bardhan:2016uhr} (derived from the available lattice results \cite{Lattice:2015rga} and from \cite{Melikhov:2000yu}), while for the $\bar{B}\rightarrow D^* \tau \nu$ decay, following \cite{Bardhan:2016uhr,Sakaki:2013bfa}, we use the heavy quark effective theory form factors based on \cite{Caprini:1997mu}. Further details about this calculation, and the analytic formulas from which \eqref{eq:allCRD} is derived, are included in \cite{Asadi:2018sym}.

Once we find the matrix elements, the differential decay rates of the $B$ meson can be calculated, and verified to be manifestly parity invariant. 

\begin{figure}
\includegraphics[scale=0.6]{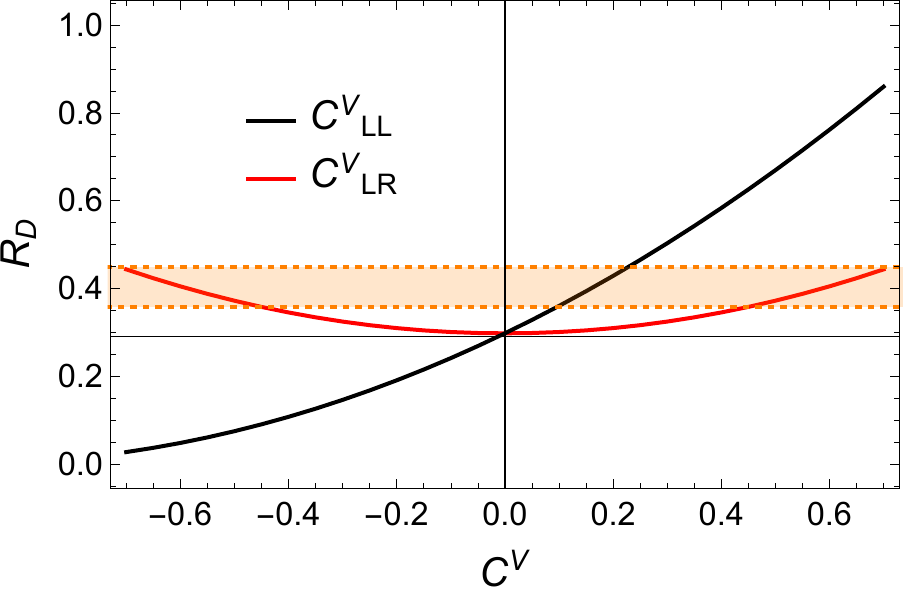}
\includegraphics[scale=0.6]{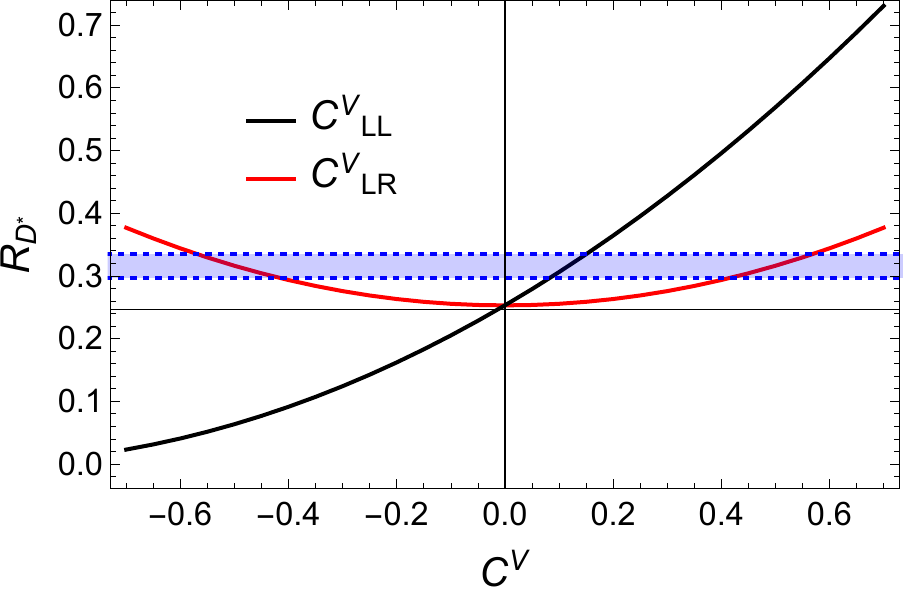}\\
\includegraphics[scale=0.6]{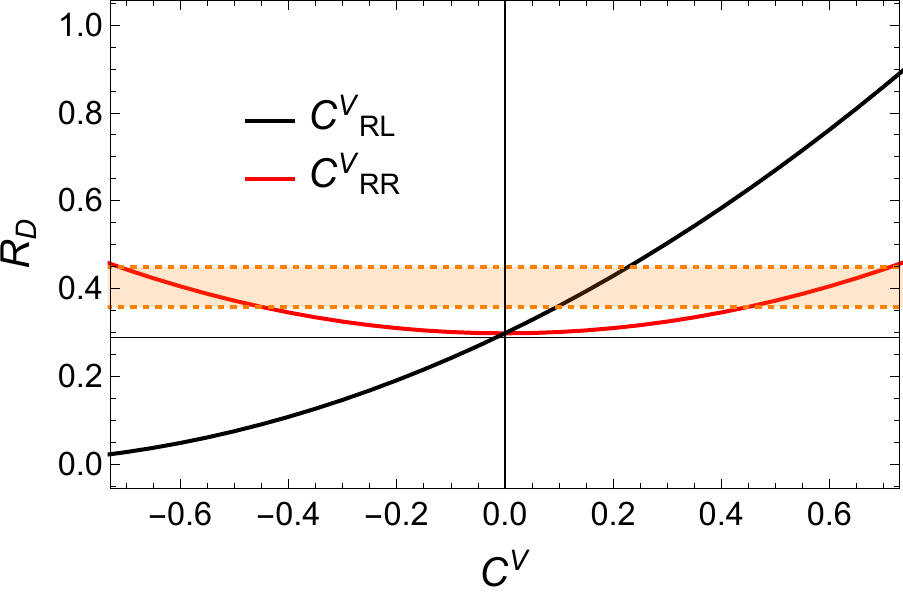}
\includegraphics[scale=0.6]{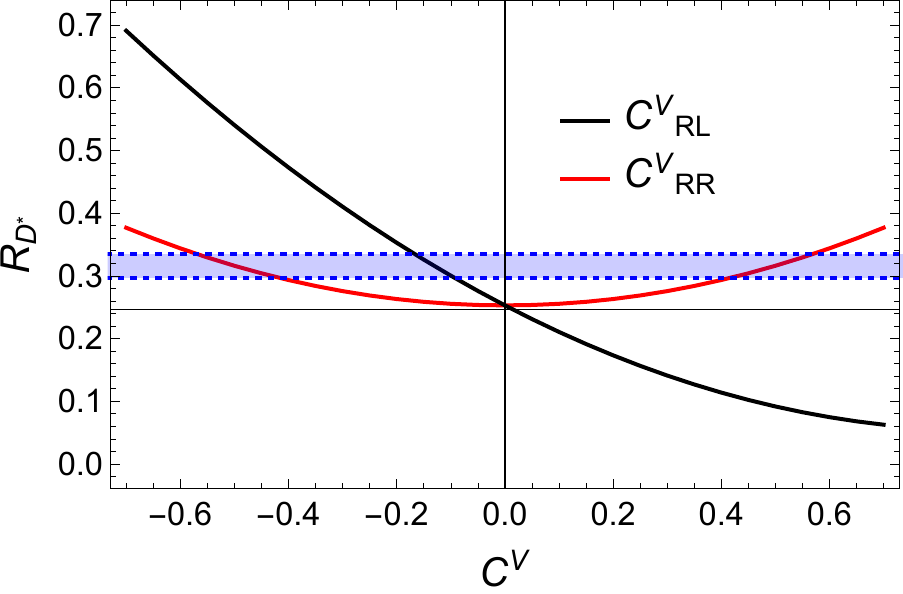}\\
\includegraphics[scale=0.6]{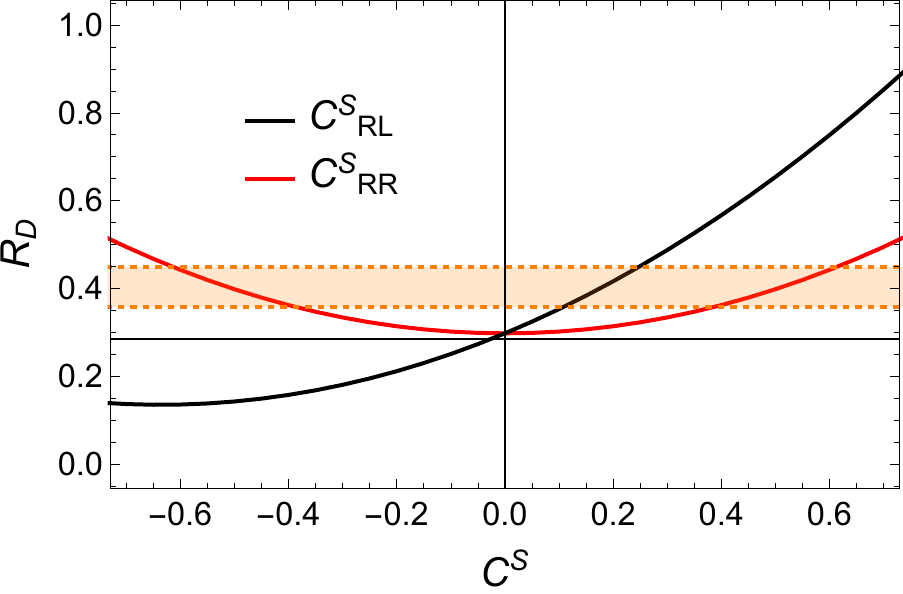}
\includegraphics[scale=0.6]{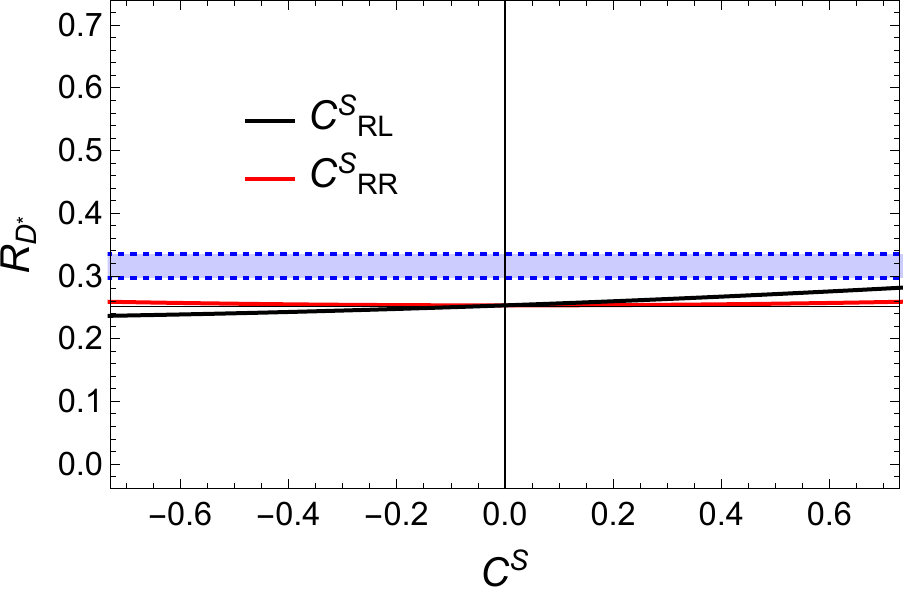}\\
\includegraphics[scale=0.6]{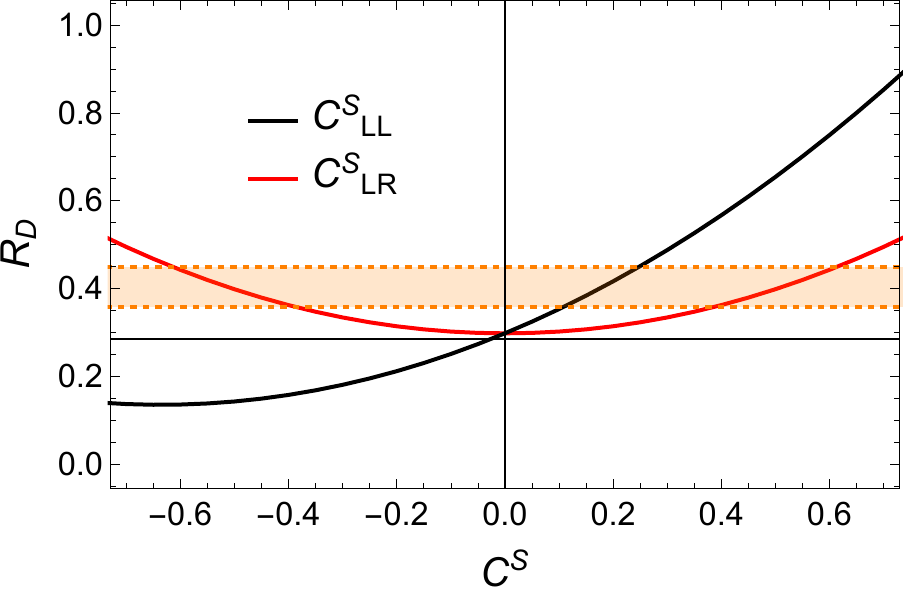}
\includegraphics[scale=0.6]{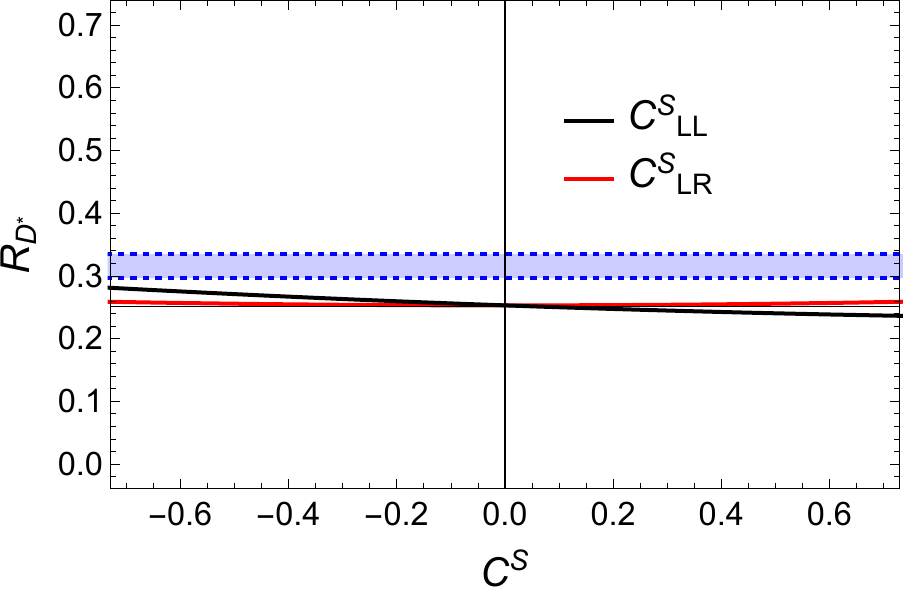}\\
\includegraphics[scale=0.6]{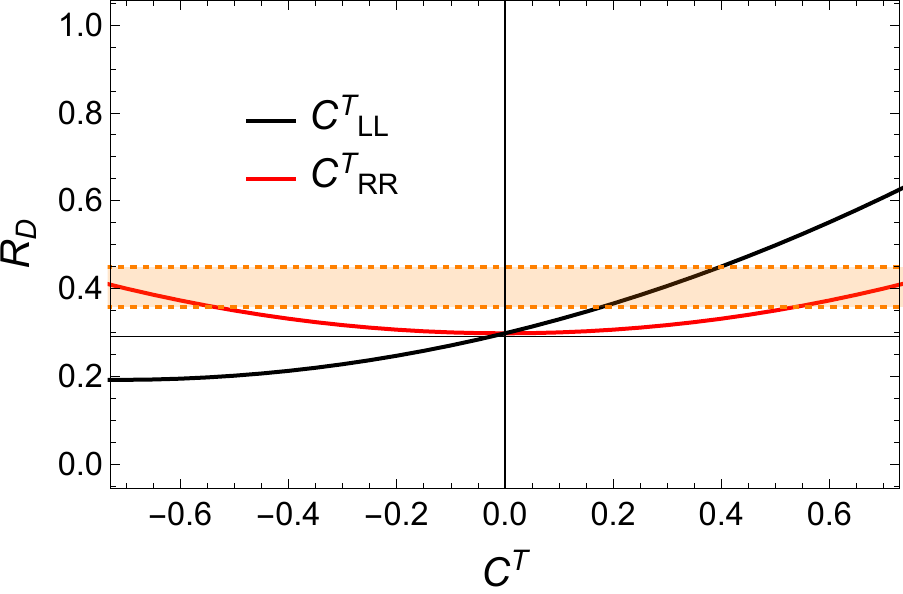}
\includegraphics[scale=0.6]{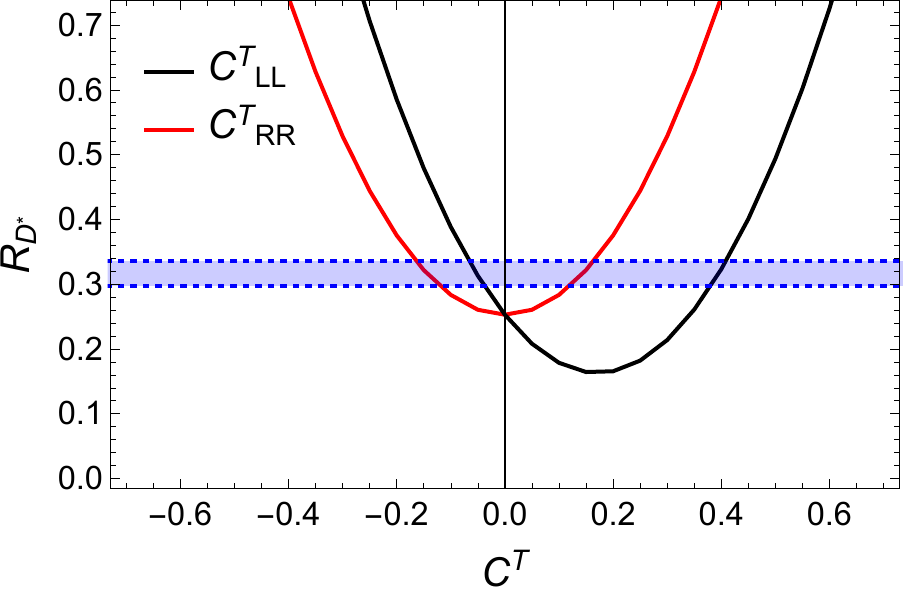}
\caption{ The dependence of $R_D$ and $R_{D*}$ on individual Wilson coefficients (with all the others being zero). The orange (blue) band indicates the $1\sigma$ band of the observed values for $R_D$ ($R_{D*}$) \cite{Amhis:2016xyh}. The qualitatively different dependence of $R_{D^{(*)}}$ on operators with left-handed neutrinos (the black lines) and those with right-handed neutrinos (the red lines) is due to the interference with the SM contribution. Each of the Wilson coefficients $C^V_{LL}$, $C^V_{RL}$, and $C^V_{RR}$ can explain both $R_D$ and $R_{D^*}$ simultaneously, thus being the most promising explanations for these anomalies.}
\label{fig:bothnurdrds}
\end{figure}

We see from \eqref{eq:allCRD} that $C^V_{LL}$, $C^V_{LR}$ and $C^V_{RR}$ are special, in that if we only turn on one of these coefficients at a time, then $R_D$ and $R_{D^{*}}$ share the same functional form. Thus a model that generates one of these coefficients will naturally explain the curious experimental fact that both  $R_D$ and $R_{D^*}$ appear to be high relative to the SM prediction by the same factor. The measured values of $R_D$ and $R_{D^*}$  can be accommodated by the other coefficients at specific points in the complex plane, but then $R_D/R_D^{SM}\approx R_{D^*}/R_{D^*}^{SM}$ would be a numerical accident, and far from natural or automatic. This is illustrated in Fig.~\ref{fig:bothnurdrds}, which shows the dependence of $R_D$ and $R_{D^*}$ on different individual Wilson coefficients (we focus in this plot on real values for simplicity). The explanation of the $R_{D^{(*)}}$ anomaly in terms of $C^V_{LL}$ is well-explored in the literature.
However, the vector operators involving right-handed neutrinos are completely unexplored and would appear, from this point of view, to be equally promising.

Specializing to just $C^V_{RR}$, the contribution of this Wilson coefficient to each anomaly is given simply by
\begin{eqnarray}
\label{eq:RDwils}
R_D & = & R_D^{SM} \left( 1+|C^V_{RR}|^2 \right), \\
\label{eq:RDswils}
R_{D^{*}} & = & R_{D^{*}}^{SM} \left(1+|C^V_{RR}|^2	\right)  .
\end{eqnarray}
We see that $C^V_{RR}$ in the range $0.4$--$0.6$ can explain both anomalies. For the rest of our phenomenological investigation we will focus on this range of this Wilson coefficient.

\section{The Model}
\label{sec:model}

In this section, we introduce our model that explains the $R_D$ and $R_{D^*}$ anomalies using a $W'$ that couples to right-handed SM fermions and a right-handed neutrino. The right-handed neutrino is assumed to be light enough ($m_{\nu_R}\lesssim 10$~keV) so that it is safe from cosmological and astrophysical bounds (see Section \ref{subsec:neutrinos}); this makes it  indistinguishable at the collider from the nearly-massless SM neutrinos in the decays of the $B$ mesons. Integrating out the $W'$ generates the $C^V_{RR}$ Wilson coefficient, capable of explaining both branching ratio measurements, as discussed in the previous section.

\begin{figure}
\includegraphics[scale=0.5]{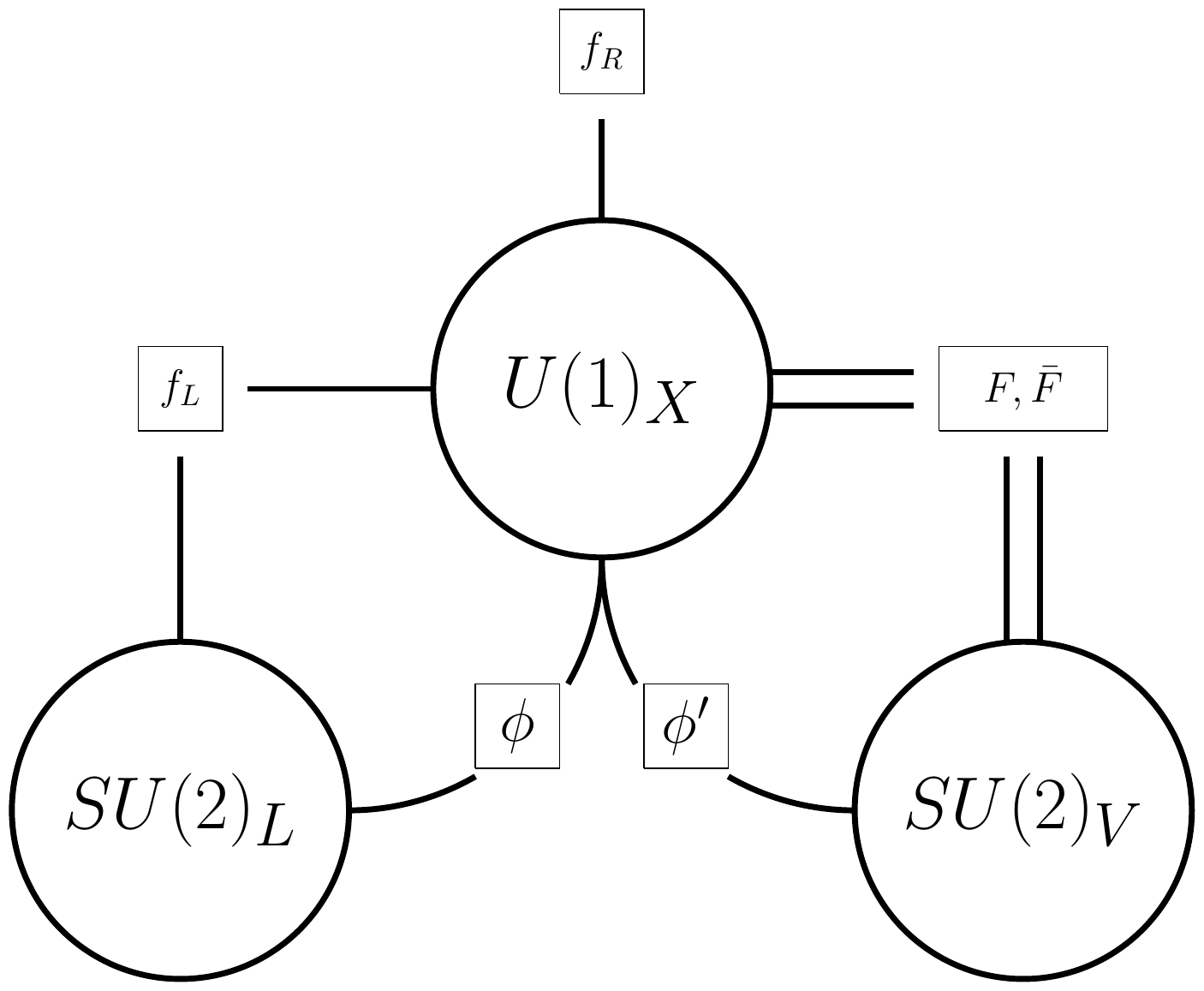}
\caption{Schematic presentation of gauge groups and matter content of our theory. We have SM-like fields charged under $SU(2)_L \times U(1)_X$ while new vector-like fermions and a new scalar $\phi'$ are charged under $SU(2)_L \times U(1)_X$. For the purpose of the anomalies of interest, only one generation of singlet $\nu_R$ is necessary. Once $\phi'$ gets a vev, one generation of the new vector-like fermions mixes with SM-like fermions through the Yukawa coupling. }
\label{fig:quivers}
\end{figure}

The field content of the model is summarized in Table \ref{tab:fields}, and a schematic presentation of the model is included in Fig.~\ref{fig:quivers}. Our model embeds hypercharge into a new $SU(2)_V\times U(1)_X$ gauge group (with gauge couplings $g_V$ and $g_X$ respectively), broken by the vev of $\langle\phi'\rangle={1\over\sqrt{2}}(0,v_V)^T$. It is useful to define the effective hypercharge coupling in our model:
\begin{equation}
g_Y^2 \equiv \frac{g_X^2 g_V^2}{g_X^2 + g_V^2}.
\label{eq:gY}
\end{equation}
After the heavy particles are integrated out, $g_L$ and $g_Y$ are identified with the SM gauge couplings, and $\phi$ is identified with the SM-like Higgs (with vev $\langle\phi\rangle={1\over\sqrt{2}}(0,v_L)^T$).

In what follows, we expand some of our equations and find the leading contribution in $v_L\ll v_V$ and $g_X$, $g_L \ll g_V$. This useful limit will simplify many of the equations that will follow. It will also prove to be a fairly good approximation in the region of the experimentally allowed parameter space capable of explaining the $B$-anomalies.

\begin{table*}[t]
\begin{tabular}{|c|c|c|c|c|c|}
\hline 
 & Generations & $SU(3)$ & $SU(2)_L$ & $SU(2)_V$ & $U(1)_X$ \\ 
\hline 
\hline
$\phi$ & 1 & 1 & 2 & 1 & 1/2 \\ 
\hline 
$q_L$ & 3 & 3 & 2 & 1 & 1/6 \\ 
\hline 
$u_R$ & 3 & 3 & 1 & 1 & 2/3 \\ 
\hline 
$d_R$ & 3 & 3 & 1 & 1 & -1/3 \\ 
\hline 
$\ell_L$ & 3 & 1 & 2 & 1 & -1/2 \\ 
\hline 
$e_R$ & 3 & 1 & 1 & 1 & -1 \\ 
\hline 
\hline
$\nu_R$ & 1 & 1 & 1 & 1 & 0 \\ 
\hline 
\hline 
$\phi'$ & 1 & 1 & 1 & 2 & 1/2 \\ 
\hline 
$Q$ & $N_V$ & 3 & 1 & 2 & 1/6 \\ 
\hline 
$L$ & $N_V$ & 1 & 1 & 2 & -1/2 \\ 
\hline 
\end{tabular} 

\caption{The field content of the model. The right-handed SM-like fermions $u_R$, $d_R$, and $e_R$ will eventually mix with the fields charged under the new gauge group $SU(2)_V$ to give rise to the actual  right-handed SM fermions. One generation of $\nu_R$, and one generation of $Q_{L/R}$, and $L_{L/R}$ mixing with SM-like fermions, are sufficient to explain the $R_D$ and $R_{D^*}$ anomalies. However, we will see in section \ref{subsec:zprimebounds} that $N_V>1$ is generally required to evade $Z'\to\tau\tau$ searches.}
\label{tab:fields}
\end{table*}

We extend the SM matter fields with a right-handed neutrino $\nu_R$ and $N_V$ generations of vector-like fermions $Q$ and $L$. In order to explain the anomalies, only one $\nu_R$ and one generation ($N_V=1$) of vector-like fermions suffices. However, we will see in Section~\ref{sec:pheno} that additional vector-like fermions (with no mixing into the SM) are required to evade direct $Z'\to\tau\tau$ searches (by enlarging the width of the $Z'$). 
 The Lagrangian of the SM is extended to\footnote{The scalar potential part of the Lagrangian is straightforward and we omit it for brevity. We can have an interaction $\bar{\nu}_R \phi \ell_L$  at tree-level as well. This operator can generate a large mass and disastrous mixing between neutrinos (see Section~\ref{sec:pheno}); hence, we must assume its Yukawa coupling is very suppressed at tree-level.}   
\bea
-\mathcal{L} &\supset M_Q \bar{Q}_L  Q_R + M_L \bar{L}_L  L_R   +m_{\nu_R} \nu_R\nu_R \\ 
&\quad +  \tilde{y}^d \bar{Q}_L \phi' b_R - \tilde{y}^u \bar{Q}_L  \phi'^{*} c_R + \tilde{y}^e \bar{L}_L \phi' \tau_R - \tilde{y}^n \bar{L}_L  \phi'^{*}  \nu_R+ \mbox{h.c.}
\label{eq:Lag}
\eea
After  $SU(2)_V\times U(1)_X$ breaking, the vector-like fermions will mix with right-handed fermions carrying SM quantum numbers. This will facilitate the interaction between the $b_R$, $c_R$, $\tau_R$ and $\nu_R$ (mediated by the $W'$ of the $SU(2)_V$) that forms the basis of our explanation of the $R_D/R_{D^*}$ anomaly. 

In the following subsections we will explore the spectrum and couplings of the model, in preparation for a detailed study of the phenomenology in section \ref{sec:pheno}.

\subsection{Gauge bosons}
\label{subsec:charged}
\label{subsec:neutrals}

The charged gauge bosons do not mix at tree-level; their spectrum is given by:
\beq\label{eq:wmasses}
m_{W}={1\over2}g_{L}v_{L},\qquad m_{W'}={1\over2}g_{V}v_{V}.
\eeq
Meanwhile, the spectrum of neutral gauge bosons is given by:
\begin{eqnarray}
\label{eq:zmass} 
m_{Z}^2  & \approx & \frac{1}{4} \left( g_L^2 + g_Y^2 \right) v_L^2  \left( 1  -  \frac{v_L^2 \varepsilon^4}{v_V^2} +	\mathcal{O} \left(\varepsilon^6 \times  \left(\frac{v_L}{v_V}\right)^4\right)	\right) , \\
\label{eq:masszplimitmass}
m_{Z'}^2 & \approx & \frac{1}{4} g_V^2 v_V^2 \left( 1 + \mathcal{O} \left(\varepsilon^4 \times \left(\frac{v_L}{v_V}\right)^2 \right) \right),
\end{eqnarray}
where $\varepsilon \equiv g_X/g_V$. 

These expressions arise from diagonalizing the following mass matrix:
\begin{equation}
\mathcal{L} \supset - \frac{1}{8}\left( 	\begin{matrix}
B & W^L_3 & W^V_3 
\end{matrix}	\right)
\left( 	\begin{matrix}
g_X^2 (v_L^2+v_V^2) & - g_L g_X v_L^2 & - g_V g_X v_V^2 \\
-  g_L g_X v_L^2 & g_L^2 v_L^2 & 0  \\
- g_V g_X v_V^2 & 0 & g_V^2 v_V^2  
\end{matrix}	\right)
\left( 	\begin{matrix}
B \\
W^L_3 \\
W^V_3 
\end{matrix}	\right)
\label{eq:zmasses}
\end{equation}
via
\begin{equation}
\left( 	\begin{matrix}
B \\
W^L_3 \\
W^V_3 
\end{matrix}	\right) \equiv \mathcal{R}^\dagger \left( 	\begin{matrix}
A \\
Z  \\
Z'
\end{matrix}	\right),
\label{eq:rotationneutrals}
\end{equation}
where, to leading order in $v_L/v_V$ and $\varepsilon$, the rotation matrix is 
\begin{equation}
\mathcal{R}^\dagger = \left( 	\begin{matrix}
\frac{g_L}{\sqrt{g_L^2+g_X^2}}&-\frac{g_X}{\sqrt{g_L^2+g_X^2}}&-\varepsilon \\
\frac{g_X}{\sqrt{g_L^2+g_X^2}}  &\frac{g_L}{\sqrt{g_L^2+g_X^2}}&0 \\
\frac{g_L}{\sqrt{g_L^2+g_X^2}} \varepsilon&-\frac{g_X}{\sqrt{g_L^2+g_X^2}}\varepsilon&1
\end{matrix}	\right) + \mathcal{O} \left(\varepsilon^2 \right).
\label{eq:neutralrotsmainlimit}
\end{equation}
In this limit we see that $W^V_3$ can be identified with the $Z'$ while the photon $A$ and $Z$ boson are a combination of $B$ and $W^L_3$ with a similar mixing pattern as in the SM. We also observe from Eq.~\eqref{eq:masszplimitmass} that $m_{Z'}$ is identical to $m_{W'}$  up to $\mathcal{O}(\varepsilon^4)$ corrections. Further details on the mass matrix and the mixing can be found in Appendix~\ref{app:details}.

\subsection{Fermion mass and mixing}
\label{subsec:fermionmassmix}

From Eq.~\eqref{eq:Lag}, after symmetry breaking, the relevant part of the fermion mass matrices can be written as
\begin{eqnarray}
\left( 	\begin{matrix}
\bar{F}_L & \bar{f}_L
\end{matrix}	\right)
\left( 	\begin{matrix}
M_F & \frac{1}{\sqrt{2}} \tilde{y}_f v_V\\
0 & m_f
\end{matrix}	\right)
\left( 	\begin{matrix}
F_R \\ 
f_R
\end{matrix}	\right),
\label{eq:upmass}
\end{eqnarray}
where $(F,f)$ refers to a paired set of a new vector-like fermion and a fermion carrying SM charges. As discussed above, to explain the anomalies without introducing FCNCs, these pairs should be  $(U,c)$, $(D,b)$, $(E,\tau)$, or $(N,\nu)$, where $(U,D)$ and $(E,L)$ come from the vector-like fermions $Q$ and $L$ respectively after $SU(2)_V$ breaking. 

Here we are implicitly working in the mass basis of the SM-like fermions, i.e.\ we imagine having already performed the CKM rotation on the left-handed SM-like matter fields, so that $m_f$ is a number, not a matrix. 

Given the structure of the mass matrix above, and the fact that the new fermion masses are much higher than SM masses, the left-handed fermions are essentially not mixed with the new vector-like fermions. As a result, for the left-handed fermions, the relationship between gauge and mass basis -- and thus the CKM matrix -- is the same as SM. 

Meanwhile, the right-handed fermions are highly mixed with the new vector-like particles.  The mixings can be parametrized by the following replacements 
\begin{eqnarray}
\left( 	\begin{matrix}
F_R \\ 
f_R
\end{matrix}	\right)\rightarrow
 \left( 	\begin{matrix}
\mathcal{U}^{f*}_{11} & \mathcal{U}^{f*}_{21}\\ 
\mathcal{U}^{f*}_{12} & \mathcal{U}^{f*}_{22}
\end{matrix}	\right)
\left( 	\begin{matrix}
F_R \\ 
f_R
\end{matrix}	\right),
\label{eq:mixingfermion} 
\end{eqnarray}
In order for the lighter mass eigenvalues to match the observed quark and lepton masses, the numerical values of $m_f$ must differ from the SM by an ${\mathcal O}(1)$ amount.

\subsection{Fermion-vector boson couplings}
\label{subsec:fvcouple}

We begin with the coupling to new gauge bosons. The mixing pattern derived in the previous section gives rise to couplings between the $W'$ gauge bosons and right-handed SM fermions: 
\beq\label{eq:WpL}
\mathcal{L}  \supset 
 \frac{g_V}{\sqrt{2}} W'_\mu \left( {\mathcal U}^{b*}_{21} {\mathcal U}^{c}_{21} \bar c_R \gamma^\mu b_R +  {\mathcal U}^{\tau *}_{21} {\mathcal U}^{\nu}_{21}\bar \nu_R\gamma^\mu \tau_R
	\right) + \mbox{h.c.} 
\eeq
The coupling to  left-handed SM fermions is highly suppressed in the large $v_V$ and $g_V$ limit, and so we neglect it in the following. After integrating out the $W'$ we generate the desired $C^V_{RR}$ operator, which can explain the $R_D/R_{D^*}$ anomaly at tree-level. In our model, the Wilson coefficient is given by:
\begin{equation}
C^V_{RR} = \frac{g_V^2 \mathcal{U}^e_{21} \mathcal{U}^\nu_{21} \mathcal{U}^d_{21} \mathcal{U}^u_{21} }{4\sqrt{2} m_{W'}^2 G_F V_{cb}} .
\label{eq:wilsonCV2R}
\end{equation}

In order to eventually study the constraints from $Z'$ resonance production in LHC, we also need the coupling of fermions to $Z'$. To leading order, the couplings of the $Z'$ to right-handed fermions will be 
\begin{eqnarray}
\label{eq:gmixedlimit}
\mathcal{L} \supset \frac{g_V}{2} Z'_\mu \left( |{\mathcal U}^{c}_{21}|^2 \bar{c}_R \gamma^\mu c_R + |{\mathcal U}^{\nu}_{21}|^2 \bar{\nu}_R \gamma^\mu \nu_R - |{\mathcal U}^{b}_{21}|^2 \bar{b}_R \gamma^\mu b_R - |{\mathcal U}^{\tau}_{21}|^2 \bar{\tau}_R \gamma^\mu \tau_R \right).
\end{eqnarray}
Even if we go beyond this leading order, we observe that the $Z'$ couplings to SM fermions are flavor diagonal and our model evades the constraining bounds from tree-level FCNCs at tree-level, as advertised. Again, the coupling of $Z'$ to the left-handed SM fermions is highly suppressed and we ignore it.

Let us now study the couplings of fermions to SM gauge bosons. These couplings will be relevant in studying EWP tests, see Section~\ref{subsec:ewp}. The coupling of $W$ to left-handed fermions has the same form as in the SM: 
\beq
\mathcal{L}\supset {1\over\sqrt{2}}g_L W^{+}_{\mu} \bar{f}_L\gamma^\mu f_L'+\mbox{h.c.}
\label{eq:Wcouplemain}
\eeq
and similarly for photons:
\beq
\mathcal{L}\supset  eQ_f A_\mu \bar{f} \gamma^\mu  f ,
\label{eq:photoncouplemain}
\eeq
where 
\beq
\label{eq:photoncharge}
e=g_L \frac{g_Y}{\sqrt{g_L^2 +g_Y^2}}, \qquad Q_f = Y + T_3^L = X + T_3^L+T_3^V.
\eeq
Finally, the coupling to the $Z$ takes the form:
\beq\label{eq:Zcouplemain}
\mathcal{L}\supset  \sqrt{g_L^2+g_Y^2} Z_\mu \Big((c^{Zf}+\delta c_L^{Zf})  \bar{f}_L\gamma^\mu f_L+(c^{Zf}+\delta c_R^{Zf})   \bar{f}_R\gamma^\mu f_R\Big),
\eeq
where
\bea
& c^{Zf}  = \left(T^L_3 - Q_f \frac{g_Y^2}{g_L^2+g_Y^2}\right)
\eea
is as in the SM, and
\bea\label{eq:deltatxt}
& \delta c_R^{Zf} \approx  Q_f \frac{v_L^2 \varepsilon^4}{v_V^2} \mp \frac{1}{2}  \frac{v_L^2 \varepsilon^2}{v_V^2}  (\mathcal{U}^f_{21})^2, \\
&\delta c_L^{Zf} \approx  (Q_f-T_3^L) \frac{v_L^2 \varepsilon^4}{v_V^2},
\eea
parametrize the deviations from the SM formulas. The minus (plus) sign in Eq.~\eqref{eq:deltatxt} is for up-type quarks (down-type quark and charged leptons); further details on these equations and couplings are included in Appendix~\ref{app:details}. 
These deviations arise either through $Z$--$Z'$ mixing (the terms that are independent of $\mathcal{U}^f_{21}$), or through fermion mixing with new vector-like fermions (the term proportional to $(\mathcal{U}^f_{21})^2$). 
Following~\cite{Efrati:2015eaa}, we will use these deviations in the couplings in our study of the EWP bounds in Section~\ref{subsec:ewp}.

\section{Phenomenology and Constraints }
\label{sec:pheno}

In this section we demonstrate that our model can generate the necessary interactions to explain the $B$-physics anomalies while evading all present constraints.

We begin by establishing the parameter space of the model.
There are six underlying parameters most relevant for our studies: the three gauge couplings $(g_L,g_X,g_V)$, the vevs $(v_L,v_V)$, and the fermion mixing parameter $\mathcal{U}_{21}$.\footnote{We assume from this point onwards that the mixing parameter is the same for all types of fermions so as to simplify our analysis.} Other parameters that we encounter in our studies can be derived from these six quantities. 

Some experimental measurements can be used to impose further relationships between these core quantities. In particular, given the precise bounds on $G_F$, $\alpha_{\rm em}$, and $m_Z$, we keep these quantities fixed at their experimentally observed values \cite{Patrignani:2016xqp}
\begin{equation}
G_F  = 1.16637 \times 10^{-5}~\mathrm{GeV}^{-2}, \hspace{0.2 in} \alpha_{\rm em} (m_Z) = 7.755 \times 10^{-3}, \hspace{0.2 in} m_Z = 91.1875~\mathrm{GeV}.
\label{eq:expvalues}
\end{equation}
We will denote the values of the gauge couplings derived from these measured quantities (assuming the SM gauge structure holds) as
\beq
\label{eq:gXothers}
\hat g_Y =0.356,\qquad \hat g_{L} = 0.650.
\eeq
We can fix $v_L$ using the relation $G_F=1/\sqrt{2}v_L^2$ (which is a tree-level relation that continues to hold in our model):
\begin{equation}
v_L = 246.2~\mathrm{GeV}.
\label{eq:vLexp}
\end{equation}
Then, we can use Eqs.~\eqref{eq:zmass} and \eqref{eq:photoncouplemain} to solve for $g_Y$ and $g_L$ in terms of the experimental values of $(\alpha_{\rm em},m_Z)$ and the other parameters of our model. To the first sub-leading order, the gauge couplings $g_Y$ and $g_L$ in our model are given by
\begin{eqnarray}
\label{eq:gY0}
g_Y & = & \hat g_Y \left( 1 - \frac{\hat g_Y^6 v_L^2}{2g_V^4 (\hat{g}_L^2-\hat{g}_Y^2) v_V^2} + \mathcal{O}\left(\varepsilon^6 \times \left(\frac{v_L}{v_V}\right)^4\right) \right) ,\\
\label{eq:gL0}
g_L & = & \hat g_{L} \left( 1 + \frac{\hat g_{L}^2 \hat g_Y^4 v_L^2}{2g_V^4v_V^2 \left( \hat g_{L}^2 - \hat g_Y^2 \right)} + \mathcal{O}\left(\varepsilon^6 \times \left(\frac{v_L}{v_V}\right)^4\right) \right) ,
\end{eqnarray}
where $\hat g_Y$ and $\hat g_{L}$ are the SM values given above. Evidently, the values of $g_Y$ and $g_L$ are shifted from their SM values by higher order corrections in $\varepsilon$ and $v_L/v_V$. 

Using the three experimentally measured quantities $(G_F,\alpha_{\rm em},m_Z)$, we have reduced the number of undetermined variables that span our parameter space to three: $(g_V,v_V,\mathcal{U}_{21})$. We choose to work in terms of the more physical parameters $(g_V,m_{W'},C^V_{RR})$, where 
\beq
\label{CV2Rallequal}
C^V_{RR}= \frac{v_L^2}{v_V^2}\frac{  (\mathcal{U}_{21})^4}{ V_{cb}}
\eeq
is derived from Eq.~\eqref{eq:wilsonCV2R} after setting all the mixing angles equal.

\subsection{Electroweak precision tests}
\label{subsec:ewp}

Our study of the EWP observables in our model closely follows the analysis in~\cite{Efrati:2015eaa}. Given the precise measurements of $G_F$, $\alpha_{\rm em}$, and $m_Z$, these quantities are fixed at their experimentally observed values. Our model can then be constrained by requiring that the NP corrections to the $W$ mass and the coupling of the $W$ and $Z$ gauge bosons to the SM fermions are within the experimental uncertainties \cite{Efrati:2015eaa}.

We saw in Eqs.~\eqref{eq:gY0}--\eqref{eq:gL0} that keeping $G_F$, $\alpha_{\rm em}$, and $m_Z$ fixed implies that $g_L$ and $g_Y$ should slightly deviate from the SM gauge couplings ($\hat{g}_L$ and $\hat{g}_Y$). This amounts to a change in $m_W$ from the SM predictions. 
Demanding the deviation in $m_W$ ($=80.385\pm0.015$~GeV) \cite{Patrignani:2016xqp} to be within the 1$\sigma$ experimental range, we find 
\begin{equation}
m_{W'} g_V \gtrsim 0.97~\mathrm{TeV}.
\label{eq:mzinequality}
\end{equation}
This is the most-constraining limit we get from EWP observables on our model.

In principle there  could be additional EWP limits coming from deviations in $W$/$Z$ couplings to fermions compared to the SM predictions. No such deviation occurs for the photon, as we have set the coupling $e$ to its experimentally observed value in Eq.~\eqref{eq:photoncouplemain}. From  Eq.~\eqref{eq:Wcouplemain}, the $W$ the coupling is $g_L$. While $g_L$ deviates from the SM value according to Eq.~\eqref{eq:gL0}, this is precisely the deviation that is being constrained by the $W$ mass measurement. The $W$ couplings to fermions do not offer any additional constraint, as they are less precisely measured than the $W$ mass. 

Finally, we consider the $Z$ couplings to fermions, shown in Eq.~\eqref{eq:Zcouplemain}. These deviations are captured by the $\delta c^{Zf}_{L,R}$ variables in Eq.~\eqref{eq:deltatxt}.\footnote{The additional deviations from $g_L\ne \hat g_L$ and $\hat g_Y\ne g_Y$ in Eqs.~\eqref{eq:gY0}--\eqref{eq:gL0} are negligible once we have satisfied the $W$ mass constraint.} The $m_W$ constraint Eq.~\eqref{eq:mzinequality} forces $v_V\gtrsim 1$~TeV, and we will see in the next subsection that $g_V\gtrsim 1$. Using these values in Eq.~\eqref{eq:deltatxt}, we find that $\delta c_R^{Zf}\lesssim 10^{-3}$ and $\delta c_L^{Zf}$ is even smaller. 

The most constraining limits on the fermion couplings are at the $({\rm few})\times 10^{-3}$ level (coming from  $\delta c_R^{Ze}$ and $\delta c_R^{Z\tau}$) \cite{Efrati:2015eaa}. Therefore, by satisfying the EWP constraint on $m_W$ and the collider bounds of the next subsection, these bounds are automatically satisfied.\footnote{As a result of a forward-backward asymmetry anomaly in LEP \cite{Efrati:2015eaa, ALEPH:2005ab}, $\delta c_R^{Zb}$ is approximately $2\sigma$ away from the SM prediction; we do not try to fit this anomaly in our model. Instead, our model predicts a very small $\delta c_R^{Zb}$, in agreement with SM predictions. According to the analysis of~\cite{Efrati:2015eaa}, the $1\sigma$ best-fit regions of some other couplings do not include the SM values either.}

\subsection{Collider Searches}
\label{subsec:zprimebounds}

Since the $W'$ and $Z'$ couple to quarks and leptons, they can be produced resonantly at the LHC. A number of different dedicated searches at LHC target such signatures \cite{Khachatryan:2016qkc, Khachatryan:2016jww, Aaboud:2016cre, Aaboud:2017sjh, Aaboud:2017efa}. In this section we study the bounds that these searches impose on our model. 

We focus on what should be the most constraining mode: resonant production of $Z'$'s that subsequently decay to $\tau^+\tau^-$ (the situation for $W'\to\tau\nu$'s should be similar). The relevant LHC searches~\cite{Khachatryan:2016qkc, Khachatryan:2016jww, Aaboud:2016cre, Aaboud:2017sjh, Aaboud:2017efa} all assumed a narrow resonance when setting their limits. We will be interested in the possibility of wide resonances (indeed, this will be necessary to evade these limits), so it is necessary to recast these searches.

Such a recast was performed for ATLAS searches of resonances decaying to high $p_T$ $\tau\tau$ final states using up to 13.2~fb$^{-1}$ of the 13 TeV dataset \cite{Aaboud:2016cre,ATLAS-CONF-2016-085,Aad:2015osa} in~\cite{Faroughy:2016osc}. This paper focused on $Z'$ models  
with mixing through left-handed SM fermions and $W'$'s and $Z'$'s that couple primarily to the 3rd generation to avoid FCNCs. As a result, the cross sections are dominated by $bb\to Z'\to\tau\tau$, and~\cite{Faroughy:2016osc} placed limits on the ratio
\begin{equation}
\eta \equiv \frac{|g_b g_\tau| v_L^2}{m_{Z'}^2},
\label{eq:kamenikcoupling0}
\end{equation}
as a function of $m_{Z'}$ and $\Gamma_{Z'}/m_{Z'}$, where $g_{b}$ ($g_{\tau}$) denotes the coupling of left-handed $b$ quarks ($\tau$ leptons) to $Z'$. For the couplings required to explain the $R_D$, $R_{D^*}$ anomaly,~\cite{Faroughy:2016osc} found that 
$\Gamma_{Z'}/m_{Z'} \gtrsim 30 \%$ was required, leading to the conclusion that perturbatively calculable $W'$ explanations of the anomaly were not viable. This is consistent with other works on $W'$ explanations of the $R_{D}/R_{D^*}$ anomaly \cite{Greljo:2015mma,Boucenna:2016qad,Boucenna:2016wpr}.

This conclusion was a consequence of assuming MFV to suppress dangerous tree-level FCNCs which, in turn, implied a $1/V_{cb}$ enhancement of the $Z'$ couplings to $bb$ relative to the $W'bc$ coupling. In our model, on the other hand, we avoid FCNCs by having the $W'$ and $Z'$ only couple to right-handed fermions. Thus our $Z'\tau\tau$ and $Z'bb$ couplings will be the same order as the $W'bc$ coupling, and the bounds from LHC searches on $Z'\to\tau\tau$ will become much less constraining. Hence we expect a smaller width to be sufficient to evade experimental bounds.

Indeed, we can see this explicitly from the formula for $\eta$ in our model. As we have
substantial $Z'cc$ couplings in addition to the coupling to $b_R$, the definition of the parameter $\eta$ of Eq.~\eqref{eq:kamenikcoupling0} must be modified to 
\beq
\eta = {v_L^2\over m_Z'^2} \,g_{R}^{Z'\tau \tau} \sqrt{(g_R^{Z'bb})^2  + \chi_c (g_R^{Z'cc})^2 } \approx V_{cb} C^V_{RR} \sqrt{1+\chi_c}+ \mathcal{O}\left(\varepsilon^3 \times \left(\frac{v_L}{v_V}\right)^3	\right),
\label{eq:kamenikcoupling}
\eeq
where $g_R^{Z'ff}$ denotes the coupling of right-handed fermion $f$ to $Z'$, Eq.~\eqref{eq:gmixedlimit}, and the second equality only contains the leading order in $v_L/v_V$ and $\varepsilon$. 
Note that the second equality also uses the assumption that all the mixing angles ${\mathcal U}_{21}^f$ are equal. Here $\chi_c$ is the ratio of the production cross-section from initial $cc$ and $bb$ states assuming identical couplings. This captures the parton distribution function (p.d.f.) enhancement from the $cc$ production channel. We obtain this ratio by simulating our model for each resonance mass using \textsc{MadGraph5\_aMC@NLO 2.5.5} \cite{Alwall:2014hca}\footnote{We use the $\mathrm{NNPDF23}\_\mathrm{lo}\_\mathrm{as}\_0130\_\mathrm{qed}$ p.d.f to calculate these production cross-sections. We also study the p.d.f and the scale uncertainties in the $Z'$ production cross-section and find less than $10\%$ error in the cross-section. This will not affect the collider bounds on our model significantly. }. Over the ranges of $m_{Z'}$ that interests us, we find $\chi_c \sim 2$--$3$. With these modifications we can apply the bounds on $\eta$ in~\cite{Faroughy:2016osc} to our model.

Interestingly, we learn that (under the simplifying assumption of equal mixings) the $R_{D}/R_{D^*}$ anomaly uniquely predicts $\eta$ and hence the rate of $bb\to Z'\to\tau\tau$ at the LHC in our model. Given the range of $C^V_{RR}$ and $\chi_c$ in our model, Eq.~\eqref{eq:kamenikcoupling} implies $\eta \in (0.026,0.048)$. For this range of $\eta$ and a generic  $Z'$ mass of $\sim 1$~TeV, Figure 4 of the recast~\cite{Faroughy:2016osc} indicates that a minimum $\Gamma_{Z'}/m_{Z'}$ of $\sim 3$--$10 \%$ is required to evade the collider bounds. 

To proceed further in applying collider limits to our model, we need a formula for $\Gamma_{Z'}$. This requires us to make a choice about the available decay channels for the $Z'$. The $Z'$ can decay into SM fermions. If kinematically allowed, it can also decay to pairs of the heavy vector-like fermions, or a single heavy fermion and a SM partner. The lower bound on new vector-like quarks is found to be above 1~TeV across a number of different searches with a variety of assumptions about decay channels \cite{ATLAS-CONF-2016-032,Sirunyan:2017ezy,Sirunyan:2017tfc,Aaboud:2017qpr,Sirunyan:2017usq,Aaboud:2017zfn,Sirunyan:2017ynj,Sirunyan:2017lzl,Sirunyan:2017pks}. We conservatively assume the new vector-like quarks are above $1.5~\mathrm{TeV}$ to evade these tight bounds. As a result of these large masses (compared to the $\sim 1$~TeV $Z'$), decays to such fermions do not contribute significantly to the $Z'$ width.

CMS has recently released a search \cite{Sirunyan:2017qkz} which significantly improves bounds on uncolored fermions. However, even these updated bounds are still far less constraining than the ones on the colored particles. The search in \cite{Sirunyan:2017qkz} targets the decay of a heavy new set of leptons into the SM charged leptons, plus $W$ and/or $Z$ gauge bosons that subsequently decay leptonicly. In particular, the $\tau$ leptons in the chain should decay leptonicly as well. The bounds from this search, however, are not that constraining for our model due to the following reasons.

\begin{itemize}

\item Given the particular mixing pattern chosen in our model, only the SM $\tau$ leptons appear in the decay chain. As indicated in \cite{Sirunyan:2017qkz}, the bounds on this tau-phillic part of the parameter space are the loosest.

\item Compared to their SM counterpart, the new gauge bosons $W'$ and/or $Z'$ in the decay chain have a lower BR into the light leptons (which is almost exclusively from the leptonic decay of a $\tau$ lepton) that further loosens the bounds on our model.

\end{itemize}

Multiplying all the $BR$s together, we get a relative suppression of the rate into light leptons compared to the model studied in \cite{Sirunyan:2017qkz}. Modifying the rates reported in 
\cite{Sirunyan:2017qkz} accordingly, the bounds on the new leptons in our model turn out far below $200$~GeV, which is the smallest mass considered in \cite{Sirunyan:2017qkz} for the new leptons. We conservatively assume all heavy leptons in our model are around $250~\mathrm{GeV}$. To enhance the $Z'$ width, we will allow there to be $N_V$ generations of new vector-like leptons (only one of which has mixing with the SM fermions).

Given the complicated expressions for the couplings, the full expression for $\Gamma_{Z'}$ is lengthy but straightforward, and we omit it here. However, a simple approximate formula (that is nevertheless fairly accurate) can be obtained if we neglect phase space suppressions and keep only the leading order expressions in $\varepsilon$ and $v_L/v_V$ (e.g.\ Eq.~\eqref{eq:gmixedlimit} for the $Z'$ couplings to SM fermions and its analogues for the heavy vector-like states):
\begin{equation}
\frac{\Gamma_{Z'}}{m_{Z'}} \approx {g_V^2\over48\pi} \left((2N_V-1) +  \mathcal{U}_{22}^4  +  \mathcal{U}_{22}^2 \mathcal{U}_{21}^2 + 4\, \mathcal{U}_{21}^4   \right).
\label{eq:Zpwidth}
\end{equation}
Using $\mathcal{U}_{21}^2+\mathcal{U}_{22}^2=1$ and  Eq.~\eqref{eq:wilsonCV2R}, we can rewrite Eq.~\eqref{eq:Zpwidth} in terms of $m_{W'}$ and $C^V_{LL}$.

The different terms in Eq.~\eqref{eq:Zpwidth} are, respectively: the decay to a pair of heavy left-handed leptons and to a pair of the heavy right-handed leptons that did not mix with the SM-like leptons (there are $N_V-1$ of these); the decay to the one pair of heavy right-handed leptons that did mix with the SM-like leptons; the decay to one heavy lepton and one SM lepton; and the decay to a pair of SM leptons and quarks. The factor of 4 in the last term is a consequence of the color factors for quarks.

\begin{figure}
\includegraphics[scale=1]{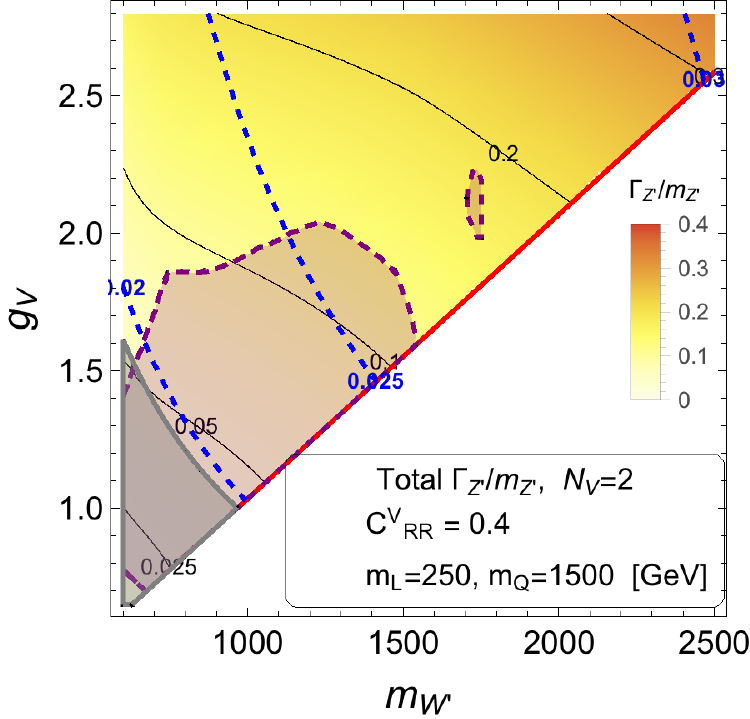}
\includegraphics[scale=1]{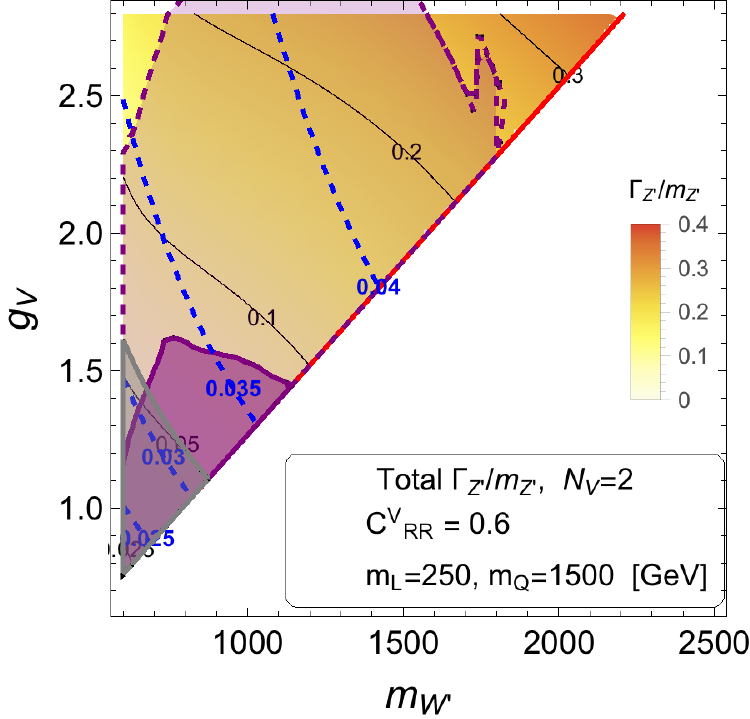}\\
\includegraphics[scale=1]{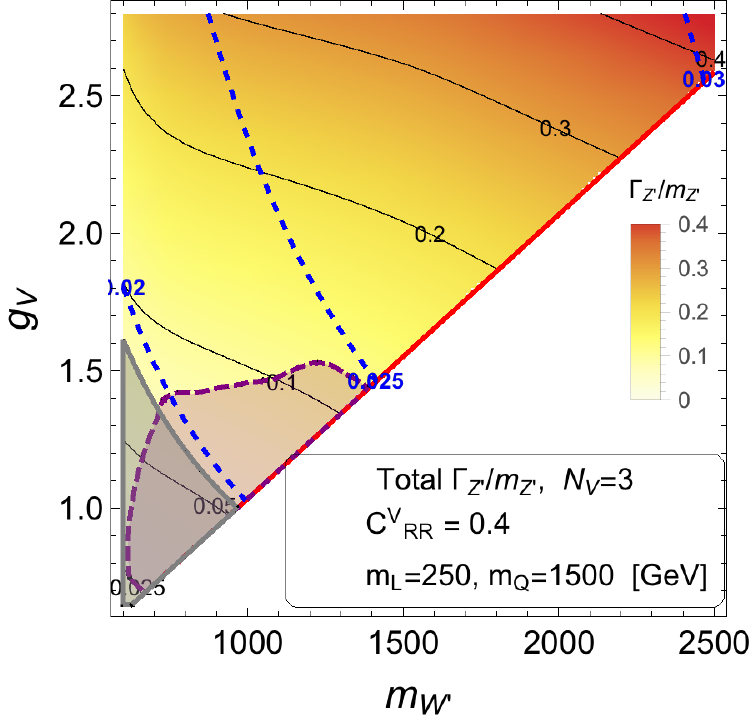}
\includegraphics[scale=1]{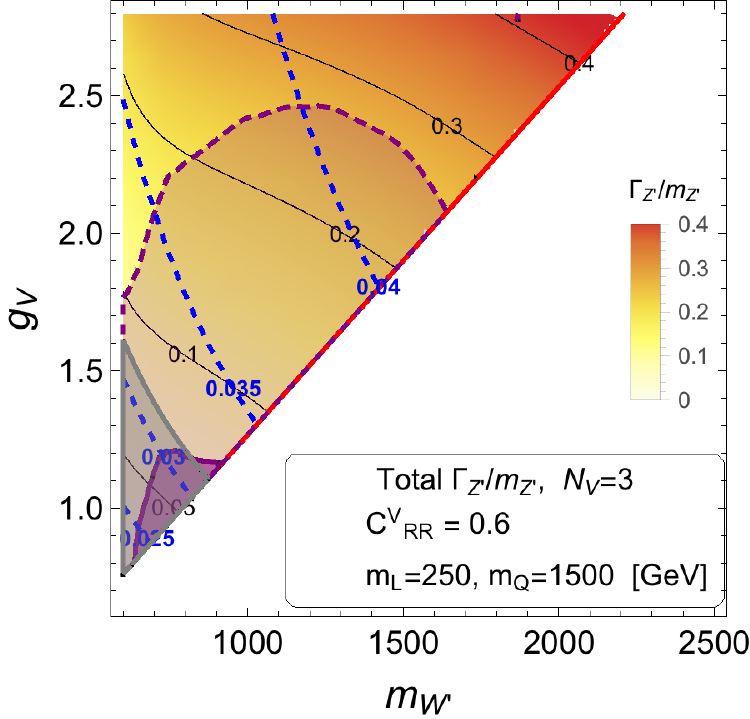}
\caption{A summary of the bounds on our model. For the left (right) plots we are assuming $C^V_{RR}=0.4$ ($C^V_{RR}=0.6$), two benchmark values that can account for the $R_{D^{(*)}}$ anomaly. Those on the top (bottom) correspond to the case $N_V=2$ ($N_V=3$) generations of new vector-like fermions, only one of which has mixing with SM fermions. We are assuming all the new leptons (quarks) have $m_L=250~\mathrm{GeV}$ ($m_Q=1500~\mathrm{GeV}$). The dashed blue curves denote the contours of constant $\eta$, while the solid black curves indicate contours of constant $\Gamma_{Z'}/m_{Z'}$. Points within the gray region have corrections to $m_W$ which are outside $1\sigma$ observed range according to~\cite{Patrignani:2016xqp}. 
(The simple inequality in Eq.~\eqref{eq:mzinequality} explains the shape of the gray lines.) 
Bounds from~\cite{Faroughy:2016osc} (obtained by recasting an older ATLAS search \cite{Aaboud:2016cre}) are indicated by the purple region (the colored region is ruled out) while a rough estimation of the bounds from a newer search \cite{Aaboud:2017sjh} are denoted by dashed purple lines. As explained in the text, adding extra generations of vector-like matter alleviates the collider bounds. }
\label{fig:result}
\end{figure}

The bounds from EWP measurements ($m_W$ more specifically) and collider searches are summarized in Fig.~\ref{fig:result}, for two representative choices of the Wilson coefficient ($C_{RR}^V=0.4$ and $C_{RR}^V=0.6$)  that can account for the $R_{D^{(*)}}$ anomaly. 
For every point below the red line, the required $\mathcal{U}_{21}$ is larger than 1, hence the indicated Wilson coefficient is not attainable in that region. It can be seen that the contours of constant $\Gamma_{Z'}/m_{Z'}$ are approximately captured by Eq.~\eqref{eq:Zpwidth}.  The contours of constant $\eta$ are also indicated; they are mostly captured by the (constant) prediction of 
Eq.~\eqref{eq:kamenikcoupling}; the small residual variation is due to variations in $\chi_c$ and higher order terms in the $\varepsilon$ and $v_L/v_V$ expansion.

One sees that for $C^V_{RR}=0.4$, $\eta$ is always small enough compared to the width ($\eta\sim 0.02-0.025$), so that there is no bound from the searches recast by~\cite{Faroughy:2016osc}. However, for $C^V_{RR}=0.6$, $\eta$ is large enough that there is a nontrivial bound. As we increase $g_V$ (holding fixed $m_{W'}$) we see that $\eta$ increases slightly (it approaches its asymptotic value given in Eq.~\eqref{eq:kamenikcoupling}), while the width increases more rapidly, as indicated in Eq.~\eqref{eq:Zpwidth} -- the coupling of $Z'$ to SM fermions becomes stronger. So moving in this direction, the limit eventually disappears. Decreasing $m_{W'}$ at fixed $g_{V}$, we see that $\eta$ decreases slightly due to subleading  corrections in $v_L/v_V$. The width decreases more significantly, in part due to phase-space suppression, but also because to hold fixed $C^V_{RR}$, we see that the fermion mixings ${\mathcal U}_{21}$ have to decrease according to Eq.~\eqref{eq:wilsonCV2R}. So we find that in this direction the limits grow stronger. The only exception is at very small $m_{W'}$, where according to the recast of Eq.~\eqref{eq:kamenikcoupling}, the limits disappear, presumably due to the kinematic thresholds of the LHC searches.

The results reported in~\cite{Faroughy:2016osc} were obtained by recasting an older ATLAS search \cite{Aaboud:2016cre}. This was updated in~\cite{Aaboud:2017sjh}. 
Given that the limits in the new search on the cross-section are improved by a factor of $\sim 3$, it is reasonable to assume that the $\eta$ bounds on the grid of Fig.~\ref{fig:result} will become a factor of $\sqrt{3}$ tighter. A crude estimate of the limits from the newer search~\cite{Aaboud:2017sjh} are shown as dashed lines in Fig.~\ref{fig:result}.

While the case $C^V_{RR}=0.6$ seems to be fairly constrained (especially with the newer search as crudely estimated in Fig.~\ref{fig:result}), we observe that for $C^V_{RR}=0.4$ the same region of the parameter space that is favored by EWP bounds is allowed by the limits on $Z'$. This region has the potential for discovery in upcoming LHC results.

\subsection{Bounds on right-handed neutrinos}
\label{subsec:neutrinos}
  
\begin{figure}
\includegraphics[scale=1]{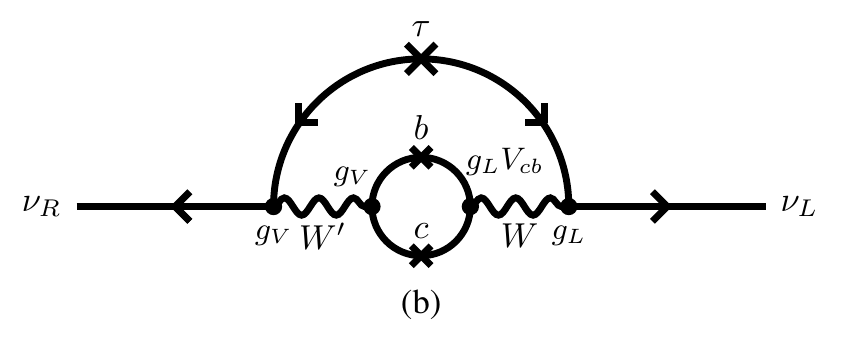}
\caption{
The only potentially dangerous loop-diagram mixing $\nu_R$ with the SM neutrinos. Other diagrams are suppressed by lack of tree-level mixing between left-handed fermions charged under different $SU(2)$ groups. Different sources of suppressions, e.g. loop factors, $V_{cb}$ suppression, and heavy mediators, will make this diagram suppressed enough so that we can evade the bounds from neutrino mixing with light-enough new neutrinos.
}
\label{fig:numixing}
\end{figure}

The right-handed neutrinos would be generated in the early Universe and so can be constrained by cosmology, assuming they are sufficiently long-lived. The lifetime depends on the right-left mixing. In a general model of right-handed neutrinos, the mass can arise from both a Dirac ($M_D$) and Majorana ($M_N$) mass term and the mixing angle $\theta$ is \cite{Kusenko:2009up}
$$|\theta| \equiv \frac{M_D}{M_N}.$$
As seen in the mass matrix in Eq.~\eqref{eq:upmass}, assuming a zero mass for left-handed neutrinos in the SM, there will be no mixing between the left-handed SM neutrinos and the new vector-like neutrinos at tree-level. Adding in the masses for the left-handed neutrinos contributes only a mixing at the level of $|\theta| \sim 10^{-20}$. However, even in the zero-mass limit for the left-handed neutrinos, there is no underlying symmetry prohibiting mixing at low energies. The dominant diagram giving rise to mixing between neutrinos is shown in Fig.~\ref{fig:numixing}. Other diagrams are significantly suppressed by the lack of tree-level mixing between $\nu_L$ and $N_L$ in our model.

To estimate the contribution of this diagram we can assume the inner loop is a mass insertion between $W$-$W'$, proportional to $m_b\times m_c$. Then we approximate the diagram and divide it by the neutrino mass to get an estimation for its contribution to the mixing $\theta$, as below
\begin{equation}
|\theta| \sim \frac{g_L^2 g_V^2 V_{cb}}{4(16\pi^2)^2} \frac{m_\tau m_b m_c}{m_{\nu_R} m_{W'} m_W}.
\label{eq:ourtheta}
\end{equation}
Inserting the range of masses and couplings in this equation suggests that our model prediction for $\theta$ is
\begin{equation}
|\theta| \sim 2 \times 10^{-5} \times \left(\frac{m_{\nu_R}}{1~\mathrm{keV}} \right)^{-1}, 
\label{eq:thetaappx}
\end{equation}

Coupling a photon to one of the charged states in the mixing diagram results in the the loop-induced decay, $\nu_R \to \nu_L \gamma$, which has a lifetime \cite{Kusenko:2009up,Bezrukov:2009th} of
\begin{equation}
\tau \approx \left(10^{30}~{\rm s}\right)\left(\frac{m_{\nu_R}}{\rm keV}\right)^{-3}.
\end{equation}
The competing tree-level $\nu_R \to 3f\bar{f}$ (where $f$ is any Standard Model fermion that couples to the $Z$ and is kinematically accessible) requires a non-zero right-left mixing angle $\theta$, and has a lifetime of
\begin{equation}
\tau \approx \left(10^{29}~{\rm s}\right)\left(\frac{m_{\nu_R}}{\rm keV}\right)^{-5}\left(\frac{\sin\theta}{2\times 10^{-5}} \right)^{-2}.
\label{eq:life3f}
\end{equation}
where $\theta$ is the mixing parameter between right- and left-handed neutrinos. As seen in Eq.~\eqref{eq:thetaappx}, the mixing angle is always small enough that we expect loop-induced $\nu_R \to \nu_L \gamma$ decays to dominate, and the lifetime is generally large compared to the age of the Universe. 

From this, we see that right-handed neutrinos below a GeV in mass are long-lived enough to be a component of dark matter (heavier $\nu_R$ are not viable replacements for the nearly-massless $\nu_L$ in the $B$-decays). Due to $Z'$-mediated pair-production and $W'$-mediated co-annihilation with $\tau$ leptons, the right-handed neutrinos would be thermally produced, in addition to any possible non-thermal production modes. These thermal processes would freeze-out around $T \sim 0.1-1$~GeV, shortly before the QCD phase transition. That is, the freeze-out occurs when the neutrinos are still relativistic. Such a dark matter candidate contributes a relic abundance directly proportional to its mass, with
\begin{equation}
\Omega h^2 \sim 10^{-1} \left[g_{*S}(T_f)\right]^{-1} \left(\frac{m_{\nu_R}}{\rm eV}\right).
\end{equation}
Assuming $T_f \sim 100$~MeV, $g_{*S} \sim 60$, and so $m_{\nu_R}$ must be less than 60~eV as to not saturate the dark matter density. This upper limit on the neutrino mass could be alleviated via non-standard cosmology, e.g. significant entropy injection \cite{Asaka:2006ek,Bezrukov:2009th}, but in any event, relativistic ``hot'' dark matter must constitute much less than 100\% of the total \cite{Aghanim:2018eyx}, pointing toward an even lighter neutrino mass. A right-handed neutrino with a mass of $\sim 10$~eV is safe from these cosmological bounds without requiring dilution.

Assuming either $m_{\nu_R} \lesssim 60$~eV or significant entropy dilution which waters down this hot contribution to dark matter, the neutrinos act as a relativistic species. These
affect the CMB power spectrum in a similar way as the SM left-handed neutrinos, shifting the time of matter-radiation equality and suppressing the power spectrum on small scales through free-streaming \cite{Osato:2016ixc, Ichikawa:2009ir}. The effect of $N$ new neutrino-like light degrees of freedom which were in thermal equilibrium with SM at some point in their history are usually quantified through the effective number of neutrinos: 
\begin{equation}
\Delta N_{\mathrm{eff}} = \left( 	\frac{g_{*}(T_\nu)}{g_{*}(T_{\nu_R})}	\right)^{4/3} N,
\label{eq:Neff}
\end{equation}
where $g_{*}(T_\nu)$ and $g_{*}(T_{\nu_R})$ are the number of relativistic degrees of freedom at the time when SM neutrinos and right-handed neutrinos decoupled, respectively. Using $g_{*}(T_{\nu_R}) \sim 80$, and $g_{*}(T_\nu) \sim 10.7$, for our model, $\Delta N_{\mathrm{eff}}\lesssim 0.07N$. The current experimental measurement is $N_{\mathrm{eff}}=3.12 \pm 0.23$ from baryon acoustic oscillations and CMB observations \cite{Ade:2015xua}. The SM prediction is $N_{\rm eff} = 3.046$; therefore, we can easily accommodate up to three light right-handed neutrinos within $1\sigma$ of the cosmological bounds. Recall that only a single species of right-handed neutrino with small mixing to the left-handed neutrinos is required in our model.

\subsection{Other bounds}
\label{subsec:otherbounds}

Besides the bounds we have already discussed, there are other potential phenomenological constraints on our model. It is straightforward to see that our model can easily evade the following bounds.

\begin{itemize}

\item \textbf{Flavor Constraints}. General mixing between the right-handed fermions could give rise to dangerous flavor-changing neutral currents. However, we have focused on a very specific mixing pattern that will suppress all the FCNCs due to $Z'$ even beyond tree-level and only couples $bc$ quarks through a $W'$, rendering the model immune to these flavor constraints. In particular, the severe bounds from neutral mesons mixing such as $K$-$\bar K$ or $B_s$-$\bar{B}_s$ mixing will not apply to our model since, due to lack of $W'$ coupling to $s$ quarks, there are no one-loop box diagrams that generate such a coupling. A recent summary of the most constraining flavor bounds for $R_{D^{(*)}}$ models can be found in \cite{Kumar:2018kmr}; we can easily see that most of these bounds are irrelevant for our model thanks to the specific fermion mixing that prohibits dangerous couplings. This pattern of couplings is \textit{ad hoc} and is solely motivated by anomalies in $bc$ interactions. It would be interesting to find a UV completion where these couplings were generated in a more natural way.

The only potential flavor constraints are those that need only a $bc$ quark flavor-changing coupling. One such observable is $B_c$ life-time. However, a symmetry similar to the one discussed in Sec.~\ref{sec:generalremarks} applies to $B_c$ life-time calculation and relates the contribution of $C^V_{RR}$ to that of $C^V_{LL}$. As the latter is not constrained (by $B_c$ life-time) for the range that explains $R_{D^{(*)}}$ \cite{Alonso:2016oyd}, neither is the former.

\item \textbf{Fermions coupling to Higgs}. Given the mixing of some SM fermions with new vector-like ones, they are effectively getting some of their mass from $\phi'$ instead of SM Higgs $\phi$. This might raise the question of how much deviation will this phenomenon give rise to in the coupling of SM fermions to $\phi$. After all, there are some constraining bounds on this deviation in the literature \cite{Sirunyan:2017khh, Khachatryan:2016vau}. However, the measured couplings are between $\phi$ and mass eigenstates and we can essentially tune the couplings of fermions charged under $SU(2)_L$ to $\phi$ such that after integrating out all the heavy degrees of freedom the effective coupling (of mass eigenstates) matches  the SM predictions. 

\item \textbf{LEP bounds}. Any vector mediator interacting with the first two generations of leptons can be subject to very stringent bounds from LEP data \cite{Alcaraz:2006mx}. However, the fermion and gauge boson mixing in our model suppresses the coupling of $Z'$ and $W'$ to the first two generations, see Appendix~\ref{app:details}, so that (except for a small part of the parameter space in Fig.~\ref{fig:result} that is already disfavored by $m_W$ limits) we automatically evade these bounds.

\end{itemize}

\section{Conclusion}
\label{sec:conclusion}

The measured ratios $R_D$ and $R_{D^*}$ are some of the largest known deviations from the predictions of the Standard Model. While they could be the result of some unknown systematic effect, no likely candidate has been identified. However, many of the proposed new physics explanations are unsatisfactory, being stringently constrained by other measurements (flavor, $B_c$ lifetime, direct collider searches, etc.)

Most of the existing explanations assume that the missing energy particle accompanying the charged tau in the decay of the $B$-meson is a left-handed SM tau neutrino. In this paper we have considered an alternative, promising hypothesis: that the anomalous measurements are the result of $b$ quarks decaying to charm and tau leptons and a new, light right-handed neutrino. After first considering all possible effective operators which alter $R_D$ and $R_{D^*}$ involving both left- and right-handed neutrinos, we focus on one particular right-handed operator that has the potential to explain both anomalies simultaneously. Further study of how different operators involving $\nu_R$ affect $R_{D^{(*)}}$ are postponed to future work \cite{Asadi:2018sym}.

This single effective operator, $\mathcal{O}^V_{RR}$, can result from integrating out a massive $W'$ that must couple to $\tau_R\nu_R$ and $b_Rc_R$. We embed this vector boson in an $SU(2)_V \times U(1)_X$ extension of the SM. However, in order to avoid an associated $Z'$ with $1/V_{cb}$ enhanced couplings to $bb$, we do not charge the SM fermions under the $SU(2)_V$. Instead, we add a generation of vector-like fermions that mix with their right-handed SM counterparts. The only coupling between the right-handed chiral quarks and leptons and the $W'$ and $Z'$ occurs through this mixing. As we show, this model can explain both the $R_D$ and $R_{D^*}$ anomalies while respecting all existing collider, cosmological, and electroweak precision bounds.

Our $W'$ model makes several concrete predictions that will be tested in the upcoming LHC data. The $W'$ and $Z'$ are close in mass, and must be below $\sim 2.5$~TeV in order to fit the anomalies with perturbative gauge couplings. In order to avoid the LHC searches for $Z'\to \tau\tau$, we require a modestly wide $Z'$ resonance ($\Gamma_{Z'} \sim 0.1m_{Z'}$). While this is safe from current limits with $30$~fb$^{-1}$ of integrated luminosity, the high-luminosity runs should be able to conclusively discover or exclude the majority of the viable parameter space. 
In addition, significant mixing with the right-handed quarks is achieved through vector-like quarks that are heavier than the existing limits ($\sim 1$~TeV), but not beyond the kinematic reach of the LHC. The width of the $Z'$ is achieved through relatively light ($\sim 250$~GeV) vector-like leptons, which are also potentially accessible at the LHC.

The $R_D$ and $R_{D^*}$ anomalies currently have no SM explanation. In considering new physics, a handful of effective operators have the desirable property of being able to explain both anomalies. The model we describe contains one such operator, along with a number of new gauge bosons, neutrinos, and vector-like quarks and leptons. The flavor anomalies, therefore, could be the first hint of a rich phenomenology hiding just beyond the current reach of the LHC, but accessible within the relatively near future.

\vskip1cm

{\bf Note added:} A similar model that also uses RH neutrinos and a $W'$ to explain the $R_{D^{(*)}}$ anomalies appeared in the contemporaneous work of \cite{Greljo:2018ogz}.

\section*{Acknowledgements}
We thank Anthony DiFranzo, Iftah Galon, Angelo Monteux, Scott Thomas, and Ryoutaro Watanabe for helpful discussions. The work of P.A. and D.S. is supported by DOE grant DE-SC0010008. M.R.B.~is supported by DOE grant DE-SC0017811.  PA thanks the Kavli Institute for Theoretical Physics for the award of a graduate visiting fellowship, provided through Simons Foundation Grant No.~216179 and Gordon and Betty Moore Foundation Grant No.~4310. This research was supported in part by the National Science Foundation under Grant No. NSF PHY17-48958.

\appendix

\section{Details of fermion-gauge boson couplings}
\label{app:details}

In this appendix we go through the details of the $Z$ boson couplings to the SM fermions in our model, which is used in our study of the EWP tests. The relevant part of the Lagrangian is 
\beq\label{eq:Lagdavidbefore}
\mathcal{L}  \supset 
\bar F\gamma^\mu (g_X X_F B_\mu +g_V T_3^V W_\mu^{3,V}) F + \bar f_L\gamma^\mu (g_X X_{f_L} B_\mu +g_L T_3^L W_\mu^{3,L}) f_L+ \bar f_R\gamma^\mu (g_X X_{f_R} B_\mu)  f_R
\eeq 
where $f_R$, $f_L$ correspond to the SM-like fermions, and $F$ to the new, heavy vector-like fermions, in the interaction basis. $B_\mu$ is the $U(1)_X$ gauge boson and $X_{F,f}$ are the $U(1)_X$ charges (given in Table~\ref{tab:fields}).

Going to the mass basis for the gauge bosons and the fermions via 
Eqs.~\eqref{eq:rotationneutrals} and \eqref{eq:mixingfermion} respectively, we obtain the couplings of $Z$ to SM fermions:
 \beq
 \mathcal{L} \supset   g_{L}^{Zf} Z_\mu 	
\bar{f}_L\gamma^\mu f_L +g_{R}^{Zf} Z_\mu 	
\bar{f}_R\gamma^\mu 
f_R
\eeq
where
\beq\label{gLZf}
 g_L^{Zf} =  g_L (T^L_3)_f \mathcal{R}^\dagger_{22} + g_X X_{f_L} \mathcal{R}^\dagger_{12}
\eeq
and
\bea\label{gRZf}
 g_R^{Zf} &= (g_V (T_3^V)_F \mathcal{R}^\dagger_{32} + g_X X_F \mathcal{R}^\dagger_{12}) |\mathcal{U}^{f}_{21}|^2 + g_X X_{f_R} \mathcal{R}^\dagger_{12} |\mathcal{U}^{f}_{22}|^2\\
 &= (g_V (T_3^V)_F \mathcal{R}^\dagger_{32} + g_X (X_F-X_{f_R}) \mathcal{R}^\dagger_{12}) |\mathcal{U}^{f}_{21}|^2 + g_X X_{f_R} \mathcal{R}^\dagger_{12} \\
  &=  (T_3^V)_F (g_V\mathcal{R}^\dagger_{32} -g_X  \mathcal{R}^\dagger_{12}) |\mathcal{U}^{f}_{21}|^2 + g_X Q_f \mathcal{R}^\dagger_{12} .
\eea
Note that for the left-handed fermions, there is essentially no mixing with the vector-like states, so the coupling to the $Z$ is relatively simple. For the right-handed fermions, we have to take into take into account mixing with the vector-like states. The choice of $F$ in Eq.~\eqref{gRZf}  is dictated by the fermion mixing. For instance, if $f_R=c_R$ ($b_R$) then $F=U$ ($D$) and $(T_3^V)_F={1\over2}$ ($-{1\over2}$). In the second line of Eq.~\eqref{gRZf}, we have used $|\mathcal{U}^{f}_{21}|^2+|\mathcal{U}^{f}_{22}|^2=1$.  In the third line we have used $Q_f=X_{f_R}=X_F+(T_3^V)_F$ for right-handed fermions and the vector-like fermions that they mix with. 

To proceed further, we require more explicit formulas for the gauge boson mixing matrix elements  $\mathcal{R}^\dagger_{i2}$. By diagonalizing the mass matrix Eq.~\eqref{eq:zmasses}, it is straightforward to verify that
\beq
\mathcal{R}^\dagger_{12}=-{g_Y\over g_X} c_\alpha s_w-{g_Y \over g_V} s_\alpha ,\qquad \mathcal{R}^\dagger_{22}=c_{\alpha} c_w,\qquad
\mathcal{R}^\dagger_{32}={g_Y\over g_X}s_\alpha - {g_Y\over g_V} c_\alpha s_w ,
\eeq
where the Weinberg angle is defined in terms of $g_Y$ and $g_L$ in the same way as the SM, and 
\begin{equation}
\tan (2\alpha) = \frac{2 v_L^2 g_X^2 \sqrt{g_V^2 g_L^2+g_X^2 g_L^2+g_V^2g_X^2}  }{-v_V^2 (g_V^2+g_X^2)^2+v_L^2 (g_V^2 g_L^2+g_X^2 g_L^2+g_V^2g_X^2-g_X^4)},
\label{eq:tan2alpha}
\end{equation}
is the effective $Z-Z'$ mixing angle (which vanishes in the $v_V\to\infty$ limit). 

Then Eq.~\eqref{gLZf} becomes
\beq\label{gLZfsimp}
 g_L^{Zf} =  \sqrt{g_L^2+g_Y^2} c_\alpha \left(  (T^L_3)_f  - Q_f  {g_Y^2\over g_L^2+g_Y^2} \right) - X_{f_L} {g_X g_Y\over g_V} s_\alpha
\eeq
and Eq.~\eqref{gRZf} becomes 
\beq
  g_R^{Zf} = -Q_f  g_Y c_\alpha s_w - Q_f {g_X g_Y  \over g_V}  s_\alpha + (T^V_3)_F s_\alpha  \sqrt{g_V^2+g_X^2}  |\mathcal{U}^{f}_{21}|^2 .
\eeq
Expanding these at large $v_V$ and $g_V$ we reproduce  Eqs.~\eqref{eq:Zcouplemain}--\eqref{eq:deltatxt} in the text.

\afterpage{\clearpage}

\bibliographystyle{utphys_modified}
\bibliography{biblio}

\end{document}